\documentclass[a4paper,times]{emsart}

\usepackage{amsthm} 

\graphicspath{{figures/}}
\usepackage{algorithm,algorithmic}

\volumeyear{xxx}
\volumenumber{xx}
\issuenumber{xx}
\journalname{Electromagnetic Science}
\startpage{x}
\DOI{xxx.xxx.xxx.xxx}
\journaltype{Review}

\title[Case studies on time-dependent Ginzburg-Landau simulations for superconducting applications]{Case studies on time-dependent Ginzburg-Landau simulations for superconducting applications}

\author{%
Cun Xue \affilnums{1}, 
Qing-Yu Wang \affilnums{2}, 
Han-Xi Ren \affilnums{3}, 
An He \affilnums{4}, and 
A V Silhanek \affilnums{5}
}

\affiliation{%
\affilnum{1}School of Mechanics, Civil Engineering and Architecture, Northwestern Polytechnical University, Xi'an 710072, China\\
\affilnum{2}School of Aeronautics, Northwestern Polytechnical University, Xi'an 710072, China \\
\affilnum{3}School of Aeronautics, Northwestern Polytechnical University, Xi'an 710072, China \\
\affilnum{4}College of science, Chang'an University, Xi'an 710064, China \\
\affilnum{5}Experimental Physics of Nanostructured Materials, Department of Physics, Universit\'{e} de Li\`{e}ge, B-4000 Sart Tilman, Belgium
}

\corrauth{An He}
\email{hean@chd.edu.cn}

\receivetime{Received \today}
\accepttime{Accepted \today}
\publishtime{Published \today}

\authorcopy{The Author(s)}

\abstract{%
The macroscopic electromagnetic properties of type II superconductors are primarily influenced by the behavior of microscopic superconducting flux quantum units. Time-dependent Ginzburg-Landau (TDGL) theory is a well-known tool for describing and examining both the statics and dynamics of these superconducting entities. They have been instrumental in replicating and elucidating numerous experimental results over the past decades. This paper provides a comprehensive overview of the progress in TDGL simulations, focusing on three key aspects of superconductor applications. The initial section delves into vortex rectification in superconductors described within the TDGL framework. We specifically highlight the superconducting diode effect achieved through asymmetric pinning landscapes and the reversible manipulation of vortex ratchets with dynamic pinning landscapes. The subsequent section reviews the achievements of TDGL simulations concerning the critical current density of superconductors, emphasizing the optimization of pinning sites, particularly vortex pinning and dynamics in polycrystalline Nb$_3$Sn with grain boundaries. The third part concentrates on numerical modeling of vortex penetration and dynamics in superconducting radio-frequency (SRF) cavities, including a discussion of superconductor–insulator–superconductor multilayer structures. In the last section, we present key findings, insights, and perspectives derived from the discussed simulations.
}


\begin{document}

\maketitle

\section{Introduction}
\label{sec1}

The hallmark of type-II superconductors is the magnetic flux penetration in the form of quantized vortices accompanied by circulating superconducting currents. The behavior of vortices determines to a large extent the macroscopic electromagnetic properties of this type of superconductors \cite{1,2}, which in turn has a direct impact on technological applications based on these materials. From an academic standpoint, this subject has captivated the research community due to the emergence of intriguing phenomena, including complex dynamic phases and transitions. The study of vortex matter entails investigating the long-range vortex-vortex and vortex-pinning interactions, on finite size samples and involving many vortices.  

The size of an individual vortex depends on whether the magnetic profile or the Cooper pair density is probed. On the one hand, probes addressing the magnetic field profile, such as scanning Hall microscopy, magnetic force microscopy, scanning squid microscopy, and many others, will reveal a vortex size on the other of the magnetic penetration depth $\lambda$ or, in case of films with thickness $t<\lambda$, the Pearl length $\Lambda=\lambda^2/t$. On the other hand, if the sensor is sensitive to the Cooper-pair density, such as in scanning tunneling microscopy, the size of the vortex will reflect the coherence length $\xi$, i.e. substantially smaller than $\lambda$. In technologically relevant superconductors, $\xi$ is on the order of few nm which implies that the macroscopic electromagnetic response of typical superconducting devices results from the interplay of millions of vortices.    

A rich diversity of phenomena associated with vortex matter have been extensively explored in the past decades. A non-exhaustive list includes the annihilation of vortices and antivortices \cite{3,4,5,6,7}, giant vortices \cite{8,9,10}, the nucleation of vortices through edge barriers \cite{11,12,13}, the interaction with artificial pinning such as antidots \cite{14,15,16,17,18} and magnetic dots \cite{19,20,21}, as well as with grain boundaries \cite{R5,R8,R35,R36,R37}. In addition, a high degree of control and manipulation of individual or clusters of vortices has been experimentally demonstrated via a localized magnetic field \cite{manipulationvortex1, manipulationvortex2,manipulationvortex3,Keren29}, 
mechanical stress \cite{manipulationvortex7}, optically \cite{manipulation}, local heating effect \cite{manipulationvortex6,manipulationvortex5}, to name a few. The diversity of approached to interact with vortices illustrates the complex multiple physical properties of these superconducting entities.


Vortices have very low mass, permitting to neglect inertial effects and consider them as moving in an overdamped medium \cite{27}. In other words, if the force acting on them is suddenly turned off, vortices stop their motion immediately without coasting down along the direction of the force. If a local oscillatory disturbance was to act on a single vortex of the lattice, no propagating vortex-lattice wave but rather a diffusing perturbation will be created. This results from the high dissipation associated to vortex motion and the associated Joule heating \cite{Bardeen-Stephen, Tinkham}. The consequent local temperature rise may spread about and weaken the effective vortex pinning strength thus favoring further vortex motion. Under certain conditions, the rapid motion of magnetic flux coupled to the associated heating effect can lead to catastrophic flux jumps and thermomagnetic instabilities responsible of unwanted magnet quenching \cite{Mints-R,wilso-new,Gilchrist,28,29}.

Exploring the behaviour of superconducting vortices at the microscopic scale is essential not only for revealing and understanding the underlying physical mechanisms behind their collective response, but also for mastering, tuning, and optimizing specific superconducting properties. The experimental tools implemented for investigating superconducting vortices comprise both indirect measurements and direct visualization. The latter benefits from the uncontested and persuasive power of an image, and frequently dodges the cumbersome layer of interpretation. Various scanning probe techniques were developed to investigate the magnetic field profile and the superconducting condensate in superconductor samples \cite{Silhanek,Bending}. Field-sensitive techniques such as scanning Hall probe microscopy \cite{b1}, scanning SQUID microscopy \cite{b2,b3,b4}, magneto-optical imaging \cite{b5}, and Bitter decoration \cite{b6} can be used to directly observe the magnetic field profile of vortices. Alternatively, one can also probe the superconducting density of the states and the supercurrents directly by scanning tunneling spectroscopy (STS) \cite{b7,b8}. Unfortunately, these techniques are typically limited to planar surfaces and rarely scalable to be adapted for the investigation of real superconducting devices such as coils or cavities. Indirect measurements, such electro-transport properties of superconducting bridges, magnetometry, ac-susceptibility, etc. typically require less sophisticated equipment although data interpretation is only possible through modeling of the sample response. In these global measurements, the details of the vortex distribution is lost and statistical averaged quantities are addressed. In other words, many microscopic states could account for the same macroscopic response.     

There are two successful approaches concerning numerical simulations of the superconducting vortex matter. One is molecular dynamics simulations based on the overdamped equation of motion, where the vortices are assumed to behave as particles, partially justified by the rigidity imposed by the flux quantum conservation of these topologically protected entities \cite{Reichhardt-n-1}. This method relies strongly on the expressions of the vortex-vortex interaction force, pinning potential caused by defects and Lorentz driving force from an external applied current. Some of the weaknesses of this approach are (i) the ignorance of possible deformations of the vortices particularly when moving at high speeds \cite{VandeVondel-2011}, or during nucleation and trapping in the vicinity of a boundary, (ii) the difficulty to implement heating effects associated to vortex motion, (iii) accounting for inhomogeneous current distributions caused by the coexistence of moving and static vortices. Despite these limitations, it has been proven to be a powerful tool able to explain matching features in superconductors with periodic pinning arrays, dynamic phases, ratchets, etc \cite{Reichhardt-n-2,Reichhardt-n-3,Reichhardt-n-4}. An alternative, and arguably more accurate method, is based on time-dependent Ginzburg-Landau (TDGL) theory. The TDGL equations describe the variations of order parameter with both time and space in superconductors \cite{43,44,45,46,47}. Although it can only be derived from the BCS theory near $T_c$ \cite{48} and for gapless superconductors, actually substantial evidence suggest that the TDGL theory is surprisingly accurate even for temperatures far below $T_c$ \cite{49,50}. Therefore, it can be used to obtain a qualitative, or even quantitative result by considering the dependence of relevant quantities on temperature \cite{43,51,52}. In addition to the vortex statics, the TDGL equations can also be used to study non-equilibrium vortex configurations \cite{53} and in the past decades, TDGL simulations have been applied to reproduce and reveal numerous experimental results \cite{54,55,56,57,58,60,61,Python,banerjee2018high,sarkar2017correlation,sarkar2016doping}.

Although the TDGL theoretical framework has a long history dating back to 1960's, TDGL equations still remains a powerful method to investigate the superconducting vortex behaviors due in part to its simplicity compared to alternative microscopic theories. In recent years, researchers have explored some technologically relevant electromagnetic properties of superconductors by means of TDGL simulations. The present work aims at showcasing recent advancements in the TDGL simulations for some applications of superconductors, namely vortex rectifications with ratchet potentials, vortex pinning and critical current of polycrystalline superconductors, and the vortex behaviors in superconducting radio-frequency (SRF) cavities with superconductor–insulator–superconductor (S-I-S) multilayer structures. These examples merely scratch the surface of TDGL's ongoing advancement towards large-scale applications. We anticipate that commercial solvers will soon incorporate TDGL into their toolkits, thereby democratizing and catalyzing the consolidation of this already thriving modeling environment.

\section{Superconducting vortex rectifiers}

Vortex rectification, vortex ratchet, or vortex diode effect generally refer to the drifting of vortices along a particular direction when they are shacked by a zero-mean symmetrical force. The resulting biased motion may arises from a pinning potential with lack of inversion symmetry. This method to induce vortex-diode effect was originally proposed by Lee \textit{et al}.\cite{c1} by considering a saw-tooth pinning potential in a superconductor in order to remove the unwanted magnetic flux trapped in it. This pioneer work fostered many theoretical and experimental efforts on the design, manufacturing, and optimization of vortex rectifiers \cite{VandeVondel-2011,c2,c3,c4,c5,c7,c8,c9,cs10}. Direction reversal of the vortex rectification was observed when asymmetric pinning sites trap vortices, due to the interaction between the trapped vortices and the mobile interstitial vortices \cite{c2}. By varying the applied magnetic field in nano-scaled asymmetric traps, de Souza Silva \textit{et al.} achieved controlled multiple reversal of the vortex rectifications \cite{c3,c4}. Through now conventional nanofababrication techniques, various asymmetric pinning landscapes, such as boomerang-shaped holes \cite{c10,c11}, triangular holes \cite{c12}, spacing-graded density of pinning sites \cite{c13}, antidots triplets \cite{c9}, conformal pinning arrays \cite{c14}, and many others were proposed in order to break the inversion symmetry of the pinning landscape in superconductors. Besides asymmetric pinning landscapes, vortex rectification can also be achieved by time-asymmetric ac current \cite{c15,c16} and asymmetric sample geometry \cite{c17,c18,c19,c20}. More recently, tunable vortex rectifications were accomplished by using an array of magnetic bars able to be configured in various spin-ice states \cite{2}.

Numerical simulations within the time-dependent Ginzburg–Landau formalism have been successfully applied to explore vortex ratchets. Berdiyorov \emph{et al.} \cite{c8} proposed a ratchet mechanism resulting from an inhomogeneous phase change in a Josephson junction. In order to realize such phase distribution, an Abrikosov vortex was strategically pinned nearby the junction. When the pinned vortex is located in the axis of symmetry of the device, vortex ratchet is absent. However, when the Abrikosov vortex is located away from the axis of symmetry, an asymmetric phase is imprinted on the Josephson junction, which leads to an asymmetric potential for the motion of a Josephson flux along the junction. 


Another interesting rectification effect happens at the sample's borders, when vortex barriers for entering and exciting the sample are different \cite{Vodolazov-n-1}. This effect has been demonstrated by introducing artificial roughness along one of the sample borders \cite{c19}. TDGL simulations provide an insight for the mechanism responsible of the ratchet \cite{c19}: the border defects enhance the current crowding, which has a significant impact on the vortex nucleation process. These effects have been often neglected when interpreting the vortex ratchets observed in arrays of asymmetric pinning potentials or, more recently, in samples exhibiting superconducting diode effect attributed to non-geometrical breaking symmetry properties \cite{Yasen-n-1}. 

Asymmetric barriers for vortex penetration are also responsible of rectification effects in ring-shaped superconductors. This situation was investigated based on TDGL equations by Jiang et al. \cite{Jiang1} for a narrow pinning-free superconducting ring. The ratchet effect is particularly prominent at low magnetic fields, weakens as magnetic field increases, and eventually exhibits a reversal ratchet signal at high magnetic fields. Interestingly, the authors show that the unusual reversal ratchet signal weakens and eventually vanishes when the frequency and ring width is increased.

An important question that was evoked in Ref. \cite{c5} concerns the characteristic spatial scale in which the broken symmetry needs to occur to induce rectification. In this work it was shown that the rectification signal is as important in small-scale as in large-scale asymmetric pinning landscapes, although the number of reversals may change significantly. In the same vein, it was later demonstrated that a superconducting film patterned with a conformal array of nanoscale holes also exhibits vortex ratchet effect \cite{Lyu}. In this case, the diode effect can be easily switched on/off by tuning the magnetic field and the rectification signal can be three orders of magnitude larger than that from a spatially-localized flux-quantum diode. An enhancement of the rectification can be achieved by lowering the temperature, which demonstrates a temperature scalable superconducting diode. The results obtained by TDGL simulations show qualitative agreement with the experimental data. 


A different way to create vortex ratchets can be realized using magnetic dipoles with in-plane magnetic moment \cite{Carneiro}. In this case, the spatial inversion symmetry is broken not by the geometry of the pinning sites but rather by the interactions between vortices and magnetic dipoles. References \cite{c4,Silhanek-2008}, unambiguously demonstrate experimentally that in-plane magnetized dipoles can indeed rectify the vortex motion. The interplay of vortices and antivortices in hybrid superconductor/ferromagnetic systems are an interesting example of the ratchet effect of binary mixtures \cite{Savelev}, where the drift direction is assisted by the interaction between the particles of different species. Some different examples can also be observed in other physical systems such as ion mixtures in cell membranes \cite{Morais-Cabral} and Abrikosov–Josephson vortex mixtures in cuprate high-Tc superconductors \cite{c15}.

The frequency range of diode effect in a superconducting microbridges of Nb${_3}$Sn has been adressed by Sara \emph{et al.} \cite{Chahid}. They documented superconducting diode rectification at frequencies up to 100 kHz. Without considering the grain boundary effect, a qualitative explanation of the diode action was obtained via finite-element modeling using TDGL equations. Based on their experimental and modeling results, they conclude that this kind of diode effect can be functional at 2–3 orders higher frequencies than 100 kHz. Dobrovolskiy \textit{et al} \cite{cs11} explored the upper frequency limits of vortex ratchet. They demonstrated that the rectifications of voltage cannot be observed for frequency exceeds 700 MHz although an ac loss response generated by vortices can be still observed at 2 GHz.

\begin{figure}[!t]
	\centering
	\includegraphics*[width=1.0\linewidth,angle=0]{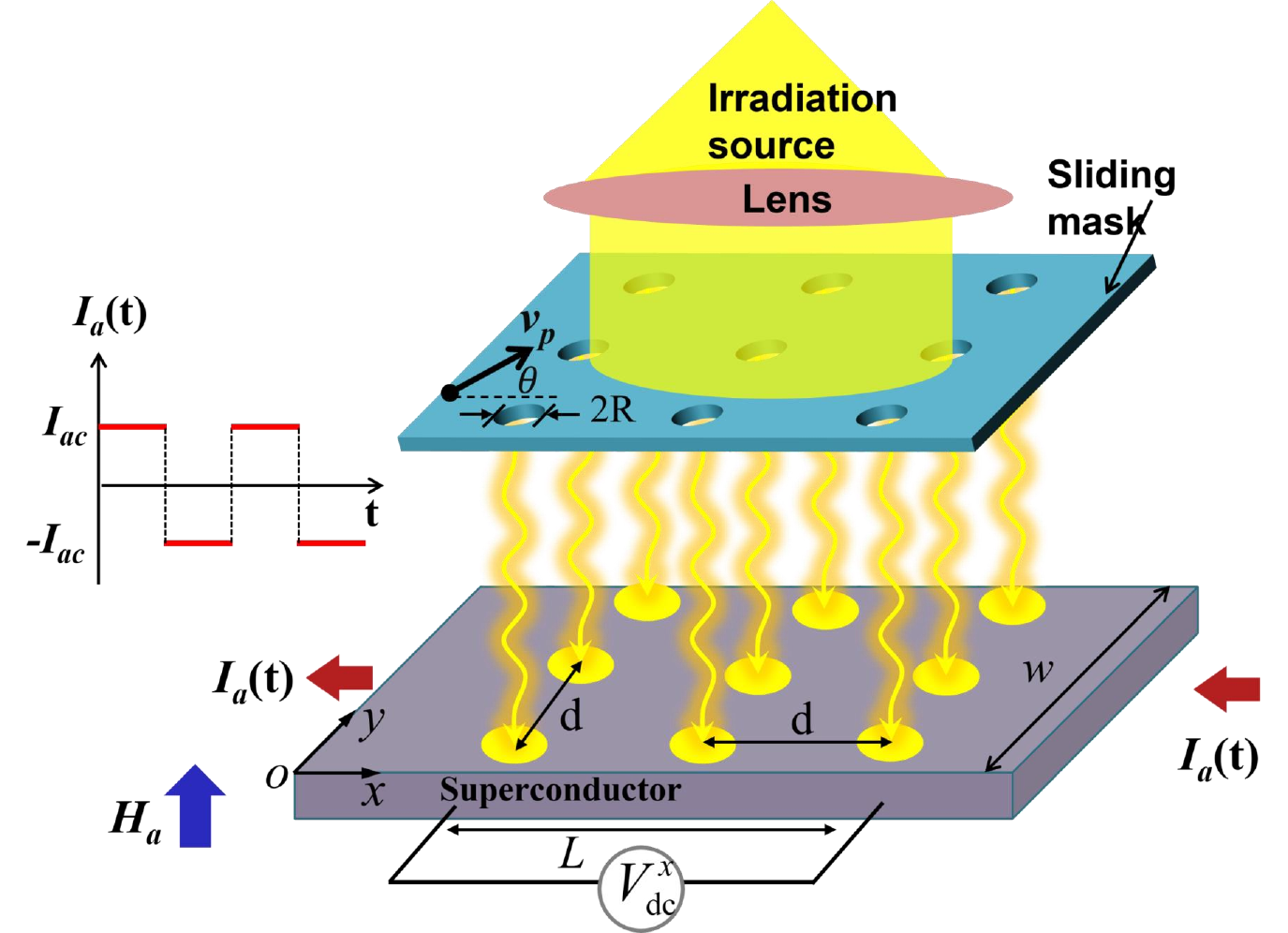}
	\caption{Schematic diagram illustrating a possible realization of a dynamic pinning potential. The sliding mask allows spatially selected filtering of irradiation which locally deplets the superconducting condensate. In the example, the layout of a sliding mask consist of a square lattice of circular holes with radius $R$ and period $d$. Besides its intensity, the dynamic pinning potential can be tuned by varying the velocity $v_{p}$ and in-plane orientation $\theta$ of its motion. The superconducting strip of width $w$ along the $y$ direction and infinite length along the $x$ direction, is applied by a square-wave ac current $I_{a}(t)$ and a perpendicular magnetic field $H_{a}$. The rectified dc voltage $V^{x}_{dc}$ is measured along the direction of applied current. Adapted from Ref. \cite{25}.}
	\label{Fig1}
\end{figure}

The above described conventional asymmetric pinning sites do not vary with time and space, offering no margin for tunability once they are manufactured. Ratchets based on magnetic landscape exhibit some degree of control, although limited to few configurations which typically require invasive procedures for switching among them \cite{Silhanek-2007}. An elegant way to achieve a higher degree of control is by introducing spatially confined and time-varying pinning potential, so called dynamic pinning centers. Recently, it has been predicted based on TDGL equations the generation of an open voltage induced by vortex dragging through a dynamic pinning landscape \cite{24}. Fig. \ref{Fig1} shows the numerical model used in simulations, which consists a superconducting strip without intrinsic pinning sites exposed to a perpendicular magnetic field $H_{a}$. The dynamic pinning landscape arranged in square array keeps moving through the strip with a certain velocity $\mathbf{v_p}$. The irradiation collimated through a sliding mask, can lead to localized hot spots or generation of quasiparticle excitations \cite{Testardi} in the superconducting film. 

The generalized TDGL equations employed to study the vortex ratchet effect with a dynamic pinning landscape are,

\begin{equation}
	\begin{aligned}
		&\frac{u}{\sqrt{1+\gamma^2|\psi|^2}}\left( \frac{\partial}{\partial t}+\frac{\gamma^2}{2}\frac{\partial|\psi|^2}{\partial t} \right)\psi \\
		&=\left( \nabla-i\textbf{A}\right) ^{2}\psi+\left( f(t,\textbf{r})-|\psi|^{2} \right)\psi,
	\end{aligned}
\end{equation}

\begin{equation}
	\frac{\partial \textbf{A}}{\partial t}={\rm{Re}} \left[ \psi^*(-i\nabla-\textbf{A})\psi \right]-\kappa^{2}\nabla\times\nabla\times\textbf{A}.
\end{equation}

The detailed normalization of the relevant physical quantities in these equations can be seen in Refs. \cite{24,25}. The parameters $u=5.79$ and $\gamma=20$ are used in the numerical simulations according to Refs. \cite{APL35,APL37,5,58}.

\begin{equation}f(t,\mathbf{r})=\begin{cases}1,|\mathbf{r}-P_0-\mathbf{v}_\mathbf{p}t|<R\\0,\quad\mathrm{otherwise}\end{cases}\end{equation}

\noindent where $P_0$ is the initial position for the centers of all pinning sites. The function $f(t,\textbf{r})$ reflects the information of dynamic pinning landscape. Periodic boundary conditions with $\psi(x,y)=\psi(x+L,y)$ and $\textbf{A}(x,y)=\textbf{A}(x+L,y)$ are used along $x-$axis. As for y-direction, $(\nabla-i\textbf{A})\psi|_{n}$=0 are used for superconductor-vacuum boundary conditions. Note that the temperature rise induced by vortex motion is normally neglected in most numerical simulations for the cases of efficient heat diffusion or slow vortex motion. This approach is justified by the fact that vortex rectification reaches an optimum performance for intermediate vortex speeds (or excitation currents) and deteriorates as vortex velocity increases \cite{Vortex2}. If, however, local temperature increase is taken into account, the natural consequence will be a decrease of the ratchet efficiency and a shift towards lower excitation currents of the optimum rectification conditions. Some numerical simulations with TDGL equations coupled with heating diffusion equation can been seen in Refs. \cite{Vodolazov11,Vodolazov12,Milo11}.

The resulting open circuit voltage $V_{dc}^x$ without applied current can be tuned by varying characteristic sizes, velocities, and densities of the dynamic pinning landscape. The edge barriers have a strong impact on the induced voltage and the vortex dynamics. In particular, significantly different features of the induced voltage are observed in the cases of dragging the vortices along the superconducting strip or perpendicular to the sample edges. Actually, the vortex dragging is similar to the well-known Giaever’s dc transformer \cite{Giaever-1965,Giaever-1966} consisting of a bilayered structure in which the vortices in a layer can be dragged by the magnetically-coupled vortices in the contiguous layer. The proposed vortex-dragging system using dynamic pinning landscapes provides new insights and a broader spectrum of possible spatio-temporal vortex manipulations. Subsequently, considering the heat dissipation equation, the heating effect on vortex dynamics of the superconducting strip using TDGL simulations was studied \cite{Chen}. The results indicate that there exist three distinct velocity ranges depending on the vortex dynamics and the induced voltage when the dynamic heat source moves along the superconducting strip. The non-linear voltage versus velocity of the dynamic heating source is similar to the case without the thermal effect \cite{24}. The dynamic pinning landscape can induce not only an open circuit voltage but also a ratchet effects under ac current \cite{25,He1}.

\begin{figure}[!t]
	\centering
	\includegraphics*[width=1.0\linewidth,angle=0]{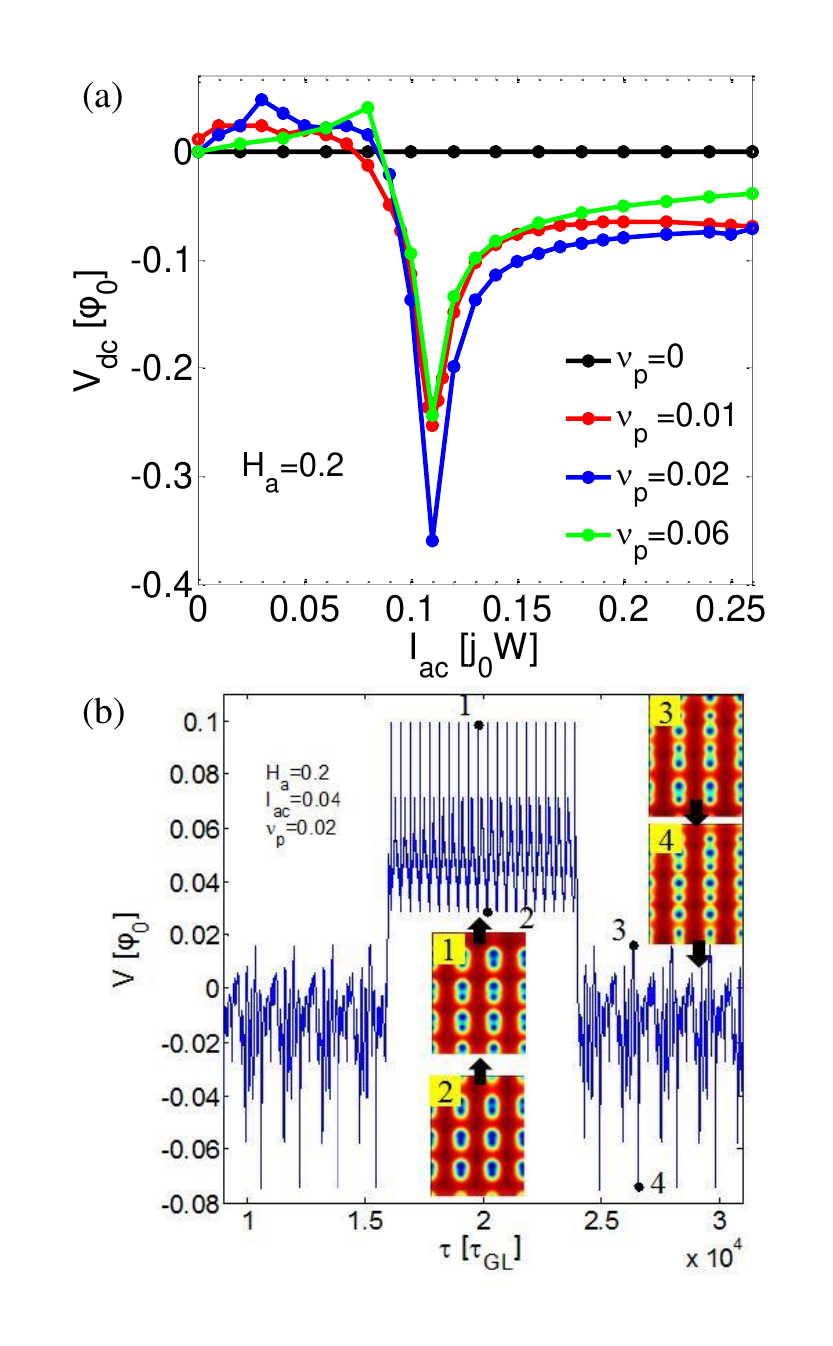}
	\caption{Rectified voltage produced by the system depicted in Fig. 1 for $\theta=90^\circ$. (a) The measured dc voltage $V_{dc}$ as a function of ac current amplitude $I_{ac}$ at a magnetic field $H_{a}$= 0.2 and temperature $T=0.8 T_c$ for several velocities $\nu_{p}$. See the related videos in Ref. \cite{25}. (b) Voltage versus time at $I_{ac}$= 0.1, $H_{a}$= 0.2 and $\nu_{p}$=0.02 as indicated in (a). The black arrows show the direction of vortex motion. Snapshops of the Cooper pair density at different instances are shown as insets. Adapted from Ref. \cite{25}.}
	\label{Fig2}
\end{figure}

\begin{figure}[!t]
	\centering
	\includegraphics*[width=0.8\linewidth,angle=0]{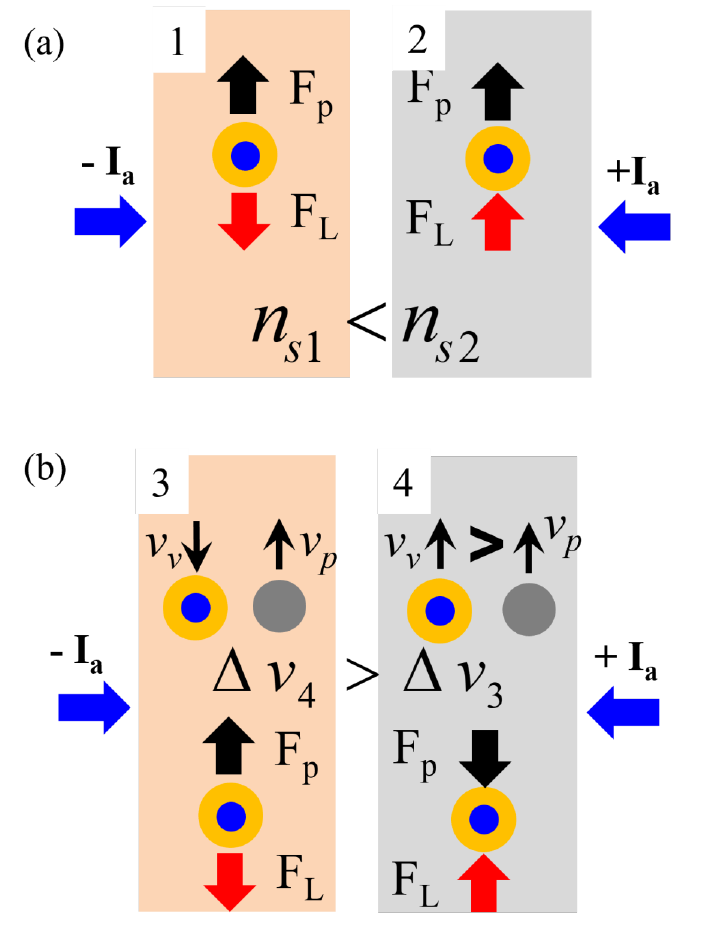}
	\caption{Schematic view of vortex rectification mechanisms with increasing ac current amplitudes. At smaller ac amplitudes, vortices move upward more easily (a), while at larger ac amplitudes, vortices move downward more easily (b). The black arrows indicate the drag force $F_p$ induced by the dynamic pinning, and the red arrows indicate the Lorentz force $F_L$ produced by $I_a$. }
	\label{Fig3}
\end{figure}

For a current-carrying superconducting strip, the superconductor is subjected to a square-wave ac current $I_{a}(t)$. The dynamic pinning potential can enhance or prevent the vortex motion depending on the ac currents \cite{VandeVondel-2010}. This leads to a vortex rectification due to the broken inversion symmetry of vortex motion. When the sliding pinning orientation is perpendicular to the direction of applied current (i.e., $\theta=90^\circ$), the Lorentz force-induced vortex motion is parallel to the direction of the sliding pinning landscape. The voltage $V^x=\int \frac{\partial \textbf{A}}{\partial t}d\textbf{l}$ is calculated between two points as indicated in Fig. ~\ref{Fig1}. The rectified dc voltage $V_{dc}$ is measured within one period of ac current. Fig. \ref{Fig2}(a) shows the evolution of the rectified dc voltage $V_{dc}$ with amplitude of ac current. There is no rectified voltage signal for a static pinning potential (with $\nu_p=0$) due to symmetric pinning effect. However, the inversion symmetry is broken when the pinning potential moves across the sample. Under this circumstance, a non-zero dc rectified voltage emerges. Note that $V_{dc}$ is finite even without any applied current due to the vortex drag caused by the sliding pinning potential. One can see a reversal of the ratchet signal with increasing ac amplitude, suggesting that vortices do not always move preferentially along the dynamic pinning potential. Schematic views of vortex rectification mechanisms at small and large ac amplitudes are shown in Fig. \ref{Fig3}. For the smaller ac amplitude, because the Lorentz force ($F_L$) and the drag force ($F_p$) act together on vortices along the same direction for +$I_a$, it is easier for vortices to moves upward than downward. The number of vortices across the sample for +$I_a$ is greater than that for -$I_a$ (see Fig. \ref{Fig3}(a)). Therefore, the voltage for the case of +$I_a$ is greater than that for the case of -$I_a$. Thus a positive rectified voltage $V_{dc}$ can be observed in the case of small $I_{ac}$. At small currents, the positive $V_{dc}$ is mainly controlled by the number of vortices across the sample because of the relative strong edge barrier effect.

With increasing ac amplitude, the rectified voltage becomes negative. This can be illustrated by the completely different $V(t)$ characteristics at time intervals with +$I_a$ and -$I_a$ in Fig. \ref{Fig2}(b). Snapshots 1 and 2 of Fig. ~\ref{Fig2}(b) show synchronous vortex motions for the case of +$I_a$, whereas snapshots 3 and 4 indicate that vortex chains move very fast along the pinning potential. This results in larger voltage for -$I_a$ than that for +$I_a$. An earlier experimental and theoretical work \cite{Formation} observed the stripe patterns in a type-II superconducting film with periodic antidots that can be stabilized by literally freezing their motion via a fast thermal quench. The arrows on the top of each snapshot in Fig.\ref{Fig2}(b) indicate the directions of instantaneous applied Lorentz force. At larger current where the Lorentz force dominates and the edge barriers are relatively weak, the pinning effect of the sliding pinning potential on the moving vortices depends on their relative velocity. At larger current the Lorentz force drives the vortices faster than the dynamic pinning. In this case the dynamic pinning impedes vortices to move upward for +$I_a$. The relative velocity $\Delta v_3$ between the moving vortices and dynamic pinning for +$I_a$ is smaller than for -$I_a$. Vortices would move more easily downward with high relative velocity as shown in Fig. \ref{Fig3}(b). This leads to a negative voltage $V_{dc}$ for larger ac amplitude.

\begin{figure}[!t]
	\centering
	\includegraphics*[width=1.0\linewidth,angle=0]{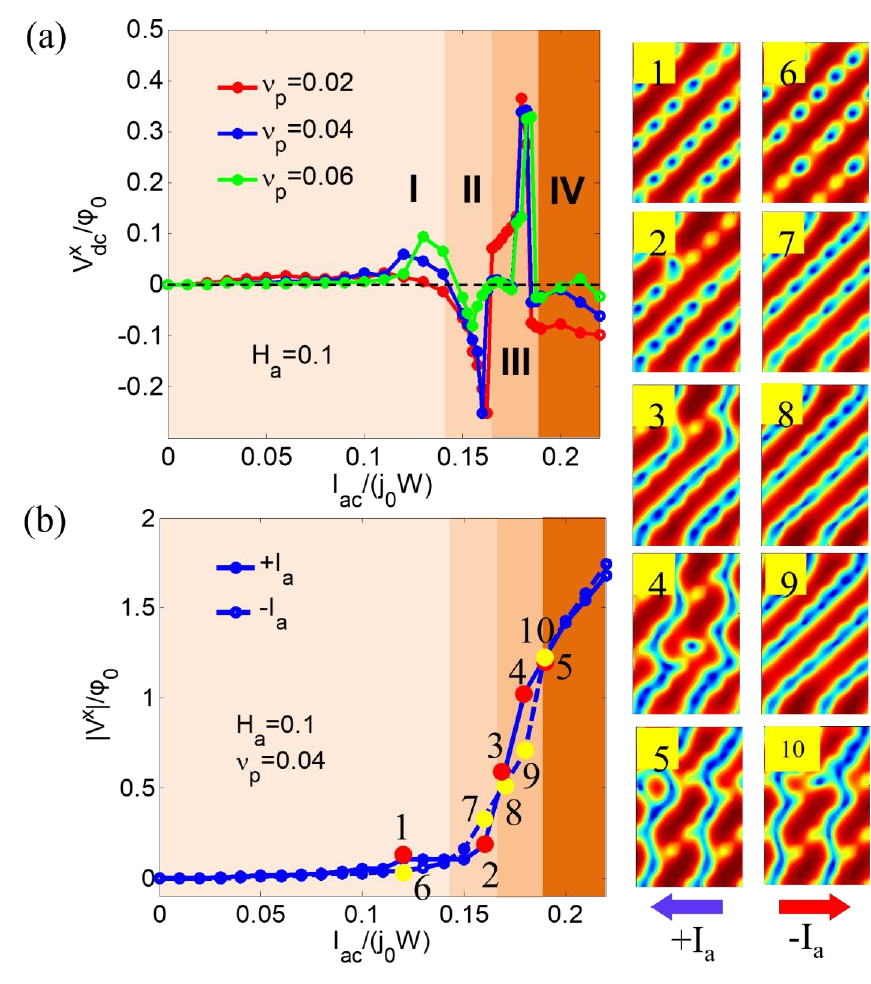}
	\caption{Average dc voltage $V^x_{dc}$ (a) as a function of ac current amplitude $I_{ac}$ at a magnetic field $H_{a}$= 0.1 for several dynamic pinning velocities $\nu_{p}$ when the sliding pinning potential moves at $\theta=45^\circ$. The absolute values of longitudinal voltage $V^x$ versus current amplitude (b). Panels 1-10 show snapshots of Cooper-pair density for $+I_a$ (left column) and $-I_a$ (right column) as indicated in the $I$-$V$ curves. The blue and red arrows indicate the directions for $+I_a$ and $-I_a$, respectively. Adapted from Ref. \cite{He1}.}
	\label{Fig4}
\end{figure}

When the sliding potential moves along $\theta=45^\circ$, the array of dynamic pinning potential moving across the sample at $\theta=45^\circ$ can induce both longitudinal and transverse ratchet effects. Fig. \ref{Fig4} shows the rectified voltages $V^{x}_{dc}$ measured along $x$ direction within one cycle of applied ac current at a magnetic field $H_{a}$=0.1. Additionally, the voltages $V^{x}$ represent the average value for $+I_a$ and $-I_a$ measured along $x$. Here we use the absolute value of $V^{x}$ both for $+I_a$ and $-I_a$. It can be found that by increasing ac current amplitude, four regions with polarity change of rectified voltage appear. Although the reversal mechanism of vortex rectifications was revealed by the change of ``easy direction" of vortex motion induced by the sliding potential discussed above, it is not sufficient to understand the unexpected multiple reversals of vortex rectification.

\begin{figure}[!t]
	\centering
	\includegraphics*[width=0.8\linewidth,angle=0]{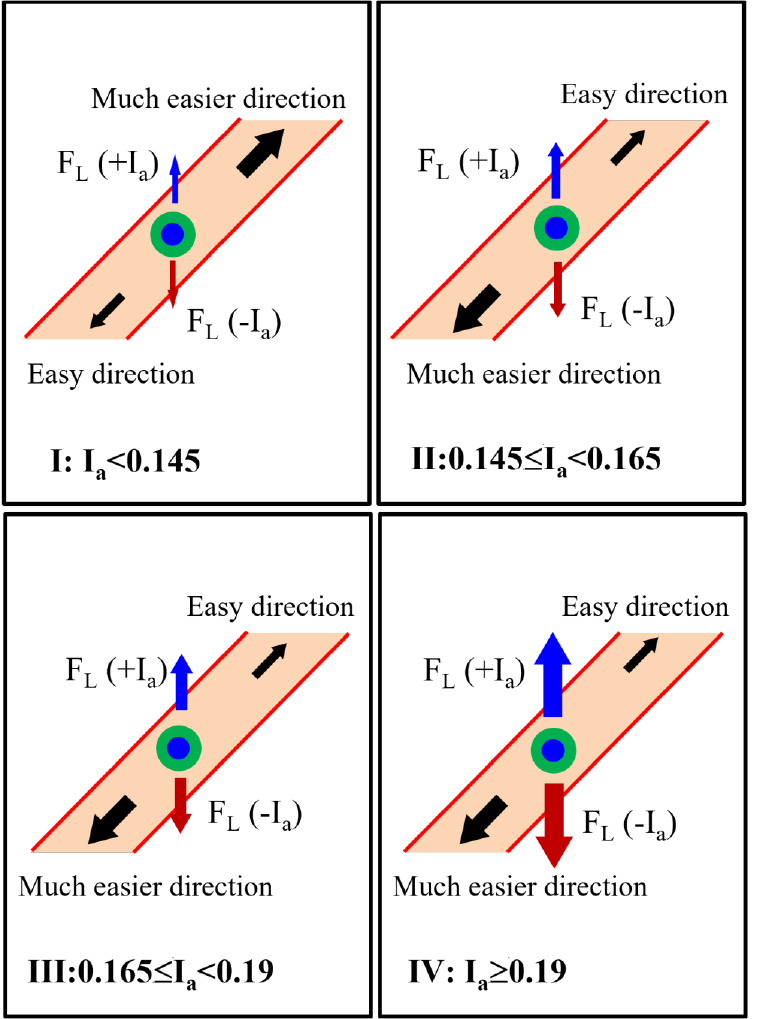}
	\caption{Schematic view of vortex rectification mechanisms with increasing ac current amplitude in which four different regimes develop. The sliding potential moves at $\theta=45^\circ$ and restricts the motion of vortices into the channel (the shaded area). The black thick arrows indicate the much easier direction for vortex motion and the thin arrows show the hard directions. The blue and red arrows indicate the Lorentz force acting on vortices for $+I_a$ and $-I_a$, respectively. Adapted from Ref. \cite{He1}.}
	\label{Fig5}
\end{figure}

We identify four distinct vortex dynamics regimes depending on the ac amplitude (see Fig. \ref{Fig5}), which can explain the physical mechanism behind the multiple reversals of rectification. In region I, at small currents, both $45^\circ$ and $-135^\circ$ are easy directions because the superconductivity along the sliding channels is suppressed. The smaller Lorentz force $F_L$ along $y$ direction cannot drive the vortices out of the sliding channels. The vortices always move along the sliding directions (panels 1 and 6 of Fig. \ref{Fig4}). On the other hand, vortices dragged by the dynamic pinning landscape along $45^\circ$ are slower than $v_p$ for $+I_a$. In other words, the $45^\circ$ direction for $+I_a$ is an easier direction than $-135^\circ$ for $-I_a$. As a consequence, one can observe a positive voltage $V^{x}_{dc}$.

With increasing ac current, in region II, the vortex velocity induced by $F_L$ is faster than $v_p$ of the sliding potential, which acts as a pinning force. The pinning strength depends on their relative velocity. The relative velocity between vortices and sliding potential for $+I_a$ is smaller than that for $-I_a$. Therefore, the pinning effect is larger for $+I_a$ than for $-I_a$. This implies that $-135^\circ$ for $-I_a$ is an easier direction. This results in a negative rectified voltage $V^{x}_{dc}$. 

In region III of Fig. \ref{Fig5}, one can observe a resurgence of a positive rectified voltage $V^{x}_{dc}$ with further increasing ac current. The sliding channels act as an easier direction, which attracts vortices moving along $-135^\circ$. In this case, the vortives cannot be driven out of the channels by $F_L$ for $-I_a$ (see panels 8-9). However, vortices can be driven our of the channels by a sufficiently strong $F_L$ for $+I_a$. Thus, one can observe some vertical vortex motion along $+y$ direction for $+I_a$ (panel 4), which leads to an important contribution to $V^{x}$ for $+I_a$, thus making the rectified voltage $V^{x}_{dc}$ to become positive again. Finally, in region IV of Fig. \ref{Fig5}, the current is so large that the Lorentz force can drive the vortices out of sliding channel. Panels 5 and 10 in Fig. \ref{Fig4} show that vortices jump out from a sliding channel and move vertically to another sliding channel for both $+I_a$ and $-I_a$. Since vortices move more 

\section{Critical current density of polycrystalline Nb$_3$Sn}

The critical current describes the  dissipation-free current-carrying capacity, which is the most important parameter for superconductors used in the manufacturing of high-field magnets. The Lorentz force makes vortices to immediately react to an applied current in a pinning-free superconductor. This motion causes energy dissipation and the presence of non-zero resistance. In this case, the movement of vortices plays a crucial role on the electromagnetic properties because it leads to an increase of the local temperature and can even lead to the catastrophic quenching of superconducting devices. Fortunately, the interaction between non-superconducting defects and vortices can immobilize the vortices by trapping them at the pinning centers. Unfortunately, the critical current density ($J_c$), which is determined by the pinning force acting on the vortices, is typically much smaller than the maximum possible, the depairing current density. Therefore, improving vortex pinning is an essential step on the quest for high critical current density \cite{Eley,Maiorov}.

TDGL theory offers a powerful tool to explore the interactions between vortices and pinning sites. With the development of computational resources, the last decades have witnessed the evolution of TDGL simulations from small samples to large-scale devices. Additionally, researchers have been able to introduce realistic pinning landscapes into TDGL simulations thus extending the research beyond regular artificial pinning sites \cite{57,60,R3,R4,R5,R8,R9}. 

Superconducting films are ideal systems to examine the effects of diverse pinning environments on the critical current. Generally, pinning centers impede the mobility of vortices and, consequently, the associated dissipation \cite{R10,R11}. Preliminary investigations have revealed that the critical current density in regularly perforated superconducting films does not exhibit a monotonic dependence with the applied magnetic field. Instead, it displays significant matching features, characterized by the presence of peaks at specific field values that correspond to a commensuration between the number of defects and vortices. Mkrtchyan and Schmidt demonstrated that the number of vortices captured by a hole depends solely on the size of the hole \cite{R12}. However, for periodic pinning arrays, the distance between holes, the temperature, and the applied field also affect the number of vortices trapped \cite{R13,R14}. Later on, experiments and numerical studies were conducted on various artificial pinning array configurations (pinscapes), such as  arrays of stripes \cite{Dobrovolskiy_3}, square arrays of antidots \cite{15,17,R18}, hexagonal (or triangular) pinning lattices \cite{R20,R21}, or honeycomb arrays \cite{R22,R23}, just to name a few. 

A forward time algorithm has been typically used in early TDGL simulations. Here, the time step depends on the grid-size and on the GL parameter $\kappa$ and thus the time step should be very small for large $\kappa$ in order to avoid divergences. A semi-implicit numerical algorithm was proposed in 2002 \cite{47}. More recently, Sadovskyy introduced GPU parallel TDGL simulations to accelerate the calculation speed \cite{57}. In this case, TDGL simulations can handle large-scale samples on a desktop computer. These authors performed TDGL simulations for a Dy-doped YBa$_2$Cu$_3$O$_{7-\delta}$ film with real pinning landscape generated by reconstructive 3D scanning-transmission-electron-microscopy tomography \cite{57}. They obtained almost the same functional dependencies and critical current values from simulations than from experiments. Furthermore, they were able to optimize the pinning landscape in order to achieve higher critical current densities. Sadovskyy used an actual Dy-doped YBa$_2$Cu$_3$O$_{7-\delta}$ film to reconstructive complex pinning landscapes and the effect of geometric confinement. Note that vortex pinning characteristics in these pinning landscapes such as inclusions, defects, and cross-sectional morphology are completely different from that in the polycrystalline superconducting materials with natural grain boundaries. In 2016, Sadovskyy \textit{et al.}  proposed a new critical current design paradigm, which aimed to predict the best defects in superconductors for targeted applications by clarifying the vortex dynamics and the associated critical current density \cite{60}, and verified it on rare earth Ba-Cu oxide coated conductors. Sadovskyy \textit{et al.} \cite{R27} utilized a large-scale numerical simulation based on TDGL equations to solve the optimization of randomly distributed metallic spherical inclusions in the vortex pinning landscape and concluded that the optimal inclusion density depends on the magnetic field. Subsequently, they conducted comprehensive research on the performance of superconductors with randomly placed spherical, elliptical, and cylindrical defects, predicting the optimal structures for pinning centers that lead to the highest achievable critical current \cite{R28}. 

In 2004, Palau \textit{et al.} \cite{palau_1} developed an inductive method which can simultaneously determine the inta- and inter-grain critical current density of YBCO coated conductors, paving the way for fully studying the relationship between grain size and vortex pinning effects. In 2006, Palau \textit{et al.} \cite{palau_2} experimentally measured the relationship between critical current ($J_c$) and temperature, magnetic field and angle of highly twinned YBa$_{2}$Cu$_{3}$O$_{7}$ thin films. They found that twin boundaries (TB) could either enhance the overall $J_c$ by serving as pinning centers or reduce $J_c$ by acting as channels for flux cutting.

It was found that for polycrystalline materials like Nb$_3$Sn, vortex pinning is primarily caused by the suppressed superconductivity at grain boundaries. Unlike the point-like pinning systems, grain boundaries (GBs) are interconnected, thus forming a mesh pattern on the plane. Experiments show that reducing the grain size can significantly improve the critical current density of Nb$_3$Sn. Xue \textit{et al.} \cite{R32} have provided a theoretical model based on the Abrikosov-Josephson (AJ) vortex distribution along a grain boundary combined with the macroscopic critical state model of high-temperature superconductors (HTS). Furthermore, Liu \textit{et al.} \cite{R33} proposed a grain boundary model based on molecular dynamics. Nevertheless, these models solely focus on investigating individual grain boundaries or a few grain boundaries rather than considering the impact of a grain boundary network on the vortex behavior. In 2021, a large-scale molecular dynamics model was developed by Xue \textit{et al.},  uncovering the role of grain boundaries on the vortex dynamics of type-II superconductors \cite{Xue_3}. By employing Voronoi diagrams they investigated the vortex dynamics of polycrystalline superconductors containing random point defects and forming grain boundary networks \cite{Xue_2}. 

In order to simulate a real polycrystalline system, a method of generating grain morphology by Voronoi mosaic has been proposed, which has become the latest approach to simulate mesoscopic polycrystalline systems \cite{R34}. The Voronoi diagram has $n$ D-dimensional polyhedra. They form an area around $n$ seed points. Each polyhedron includes all points closer to its seed point than to any other point in the region. Physically speaking, this is equivalent to a polycrystalline material in which all grains nucleate simultaneously and grow uniformly at the same rate \cite{R35}. The structures generated by this method are in agreement with the actual size of polycrystalline grains and their nearest-neighbor distribution \cite{R35} and  have been widely used in polycrystal simulations, including those that consider the impact of grain size on grain boundary resistivity \cite{R36}. Blair and Hampshire used this method to simulate a small-sized two-dimensional (2D) polycrystalline system \cite{R37}, and obtained as preliminary conclusion that grain boundaries affect the critical current properties. Although the simulation captured the actual polycrystalline system, it still failed to achieve the macroscopic statistical effect. In order to address this issue, it is necessary to combine Voronoi mosaic and GPU parallel computations for simulating a sufficiently large sample.

\begin{figure} [tb]
	\centering 
	\includegraphics*[width=0.9\linewidth,angle=0]{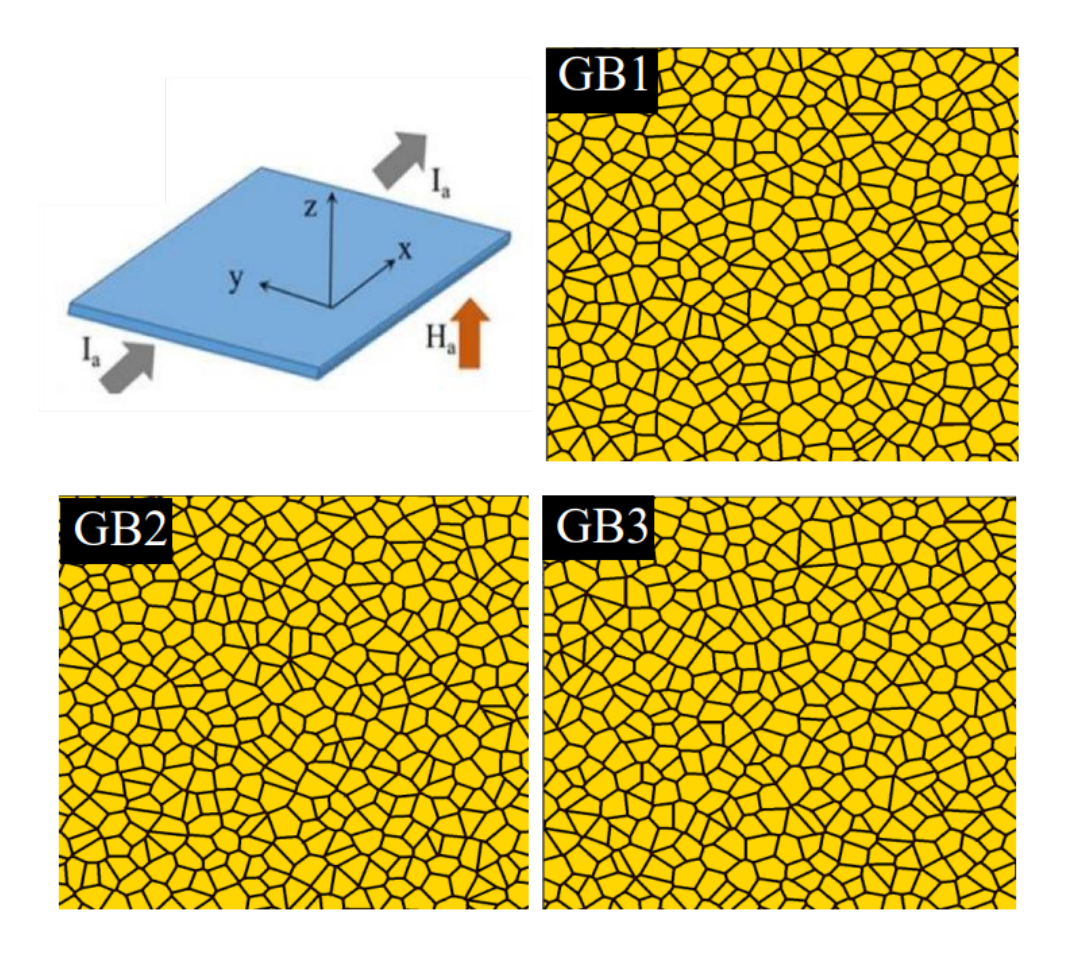}
	\caption{Schematic diagram of numerical model for a current-carrying superconducting film in a vertical magnetic field and three different patterns of GBs with a constant grain size of thirty-eight (approximately  115 nm in SI unit) produced by voronoi polygons. Adapted from Ref. \cite{R8}.}
	\label{Figmodel}
\end{figure}

\begin{figure} [tb]
	\centering 
	\includegraphics*[width=1.06\linewidth,angle=0]{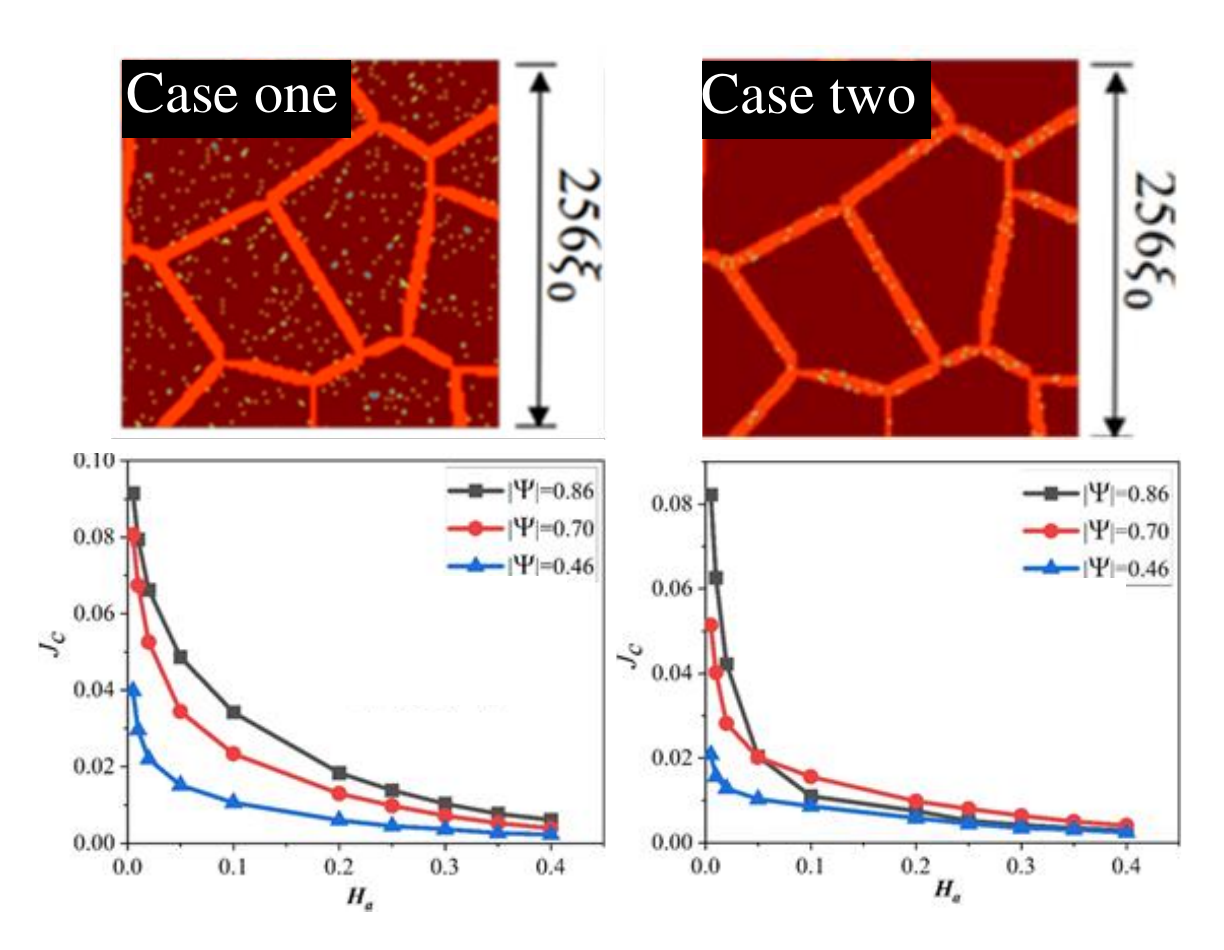}
	\caption{The variations of maximal ($J_c$) with magnetic field by varying the density of artificial pinning sites for two cases where artificial pinning sites are randomly distributed only inside the crystalline grains (case 1), and random artificial pinning sites only distributed on GBs (case 2).}
	\label{Effect}
\end{figure}

For a current-carrying polycrystalline superconducting thin film exposed to a magnetic field as shown in Fig.~\ref{Figmodel}, the random artificial pinning sites and grain boundaries can be introduced into TDGL equations as follows,

\begin{equation}
	\begin{split}
		u(\partial_{t}+i\tilde{\mu})\big[b(\mathbf{r})\tilde{\psi}\big]= & (\nabla-\mathrm{i}A)^2\big[b(\mathbf{r})\tilde{\psi}\big]+\\  & \epsilon(\mathbf{r})b(\mathbf{r})\tilde{\psi}-b(\mathbf{r})\big|\tilde{\psi}\big|^2\tilde{\psi}
	\end{split}
\end{equation}

\begin{equation}\Delta\tilde{\mu}=\nabla\mathrm{Im}\left\lbrace b\left(r\right)\tilde{\psi}^{*}\left(\nabla-\mathrm{i}\widetilde{A}\right)\left\lbrack b\left(r\right)\tilde{\psi}\right\rbrack\right\rbrace\end{equation}

\noindent where $\epsilon(\textbf{r})$ can be used to tune the superconductivity of the grain boundary by decreasing $T_c$; $b(\textbf{r})$ equals to 1 in the superconducting regions and 0 elsewhere. Note that both $\epsilon(\textbf{r})$ and $b(\textbf{r})$ should be periodic in order to satisfy the periodic boundary condition.

The grain structure in Fig. ~\ref{Figmodel} is characterized by an adjustable average grain size. When the mean grain size is constant, the size distribution follows a Gaussian distribution. In the numerical simulation, the pinning can vary with the suppression of superconductivity at the grain boundaries. Additionally, artificial pinning sites are randomly introduced in the simulated region. 

As shown in the Fig.~\ref{Figmodel} \cite{R8}, a number of different GB patterns are created by Voronoi diagram, and these different GB configurations exhibit a consistent average grain size. The critical current densities of different GB patterns are calculated by using two different average Cooper pair densities. The simulation results show that for a given $|\psi|_{GB}$, the $J_c$ are nearly identical. This demonstrates that the modeled region is large enough, and the simulated $J_c$ is a statistically reliable result, which is not affected by the randomly generated GB configurations.

The relationship between the critical current density and the magnetic field, obtained using the aforementioned approach, is compared with the widely acknowledged scaling law \cite{74} and a modified scaling law \cite{75} that have been substantiated through numerous experimental validations \cite{76}. It illustrates that the simulated results agree well with the scaling law and the modified scaling law from experiments (see Fig. 10 in Ref.~\cite{R8}). Those authors investigated the critical current density in polycrystalline systems \cite{R9, R8} and demonstrated that the relationship between critical current density and the superconductivity at the grain boundaries is non-monotonic \cite{R8}. Furthermore, the optimal order parameter of GBs to achieve maximum critical current depends on the applied magnetic field. For higher magnetic fields (10-16 T), it was found that the optimal superconductivity (order parameter) of GBs is approximately 0.8-0.9.

In addition, the effect of grain size on critical current density was also explored by numerical simulations \cite{R8}.  Refining the grains enhances $J_c$ under weak magnetic fields ($H_a=0.005$). However, in a strong magnetic field, $J_c$ varies complexly.  At $H_a=0.5$, smaller grain sizes only raise critical current density if the superconductivity at grain boundaries is slightly suppressed. But, if superconductivity at GBs is greatly suppressed, then critical current density drops with smaller grains (see Fig. 7 and Fig. 8 in Ref.~\cite{R8}).

In addition to the aforementioned investigations on polycrystalline superconducting materials, we have performed further numerical simulations by introducing artificial pinning sites. Here we consider two different cases, i.e., (i) the artificial pinning sites only randomly distributed inside crystalline grains and (ii) random artificial pinning sites only distributed on the GBs. Fig.~\ref{Effect} shows the maximal current density by varying the density of artifical pinning sites at various applied magnetic fields. Comparing these two cases, one can see that the critical current decreases more rapidly in the last scenario than in the other two cases. This indicates that artificial pinning sites should be introduced inside crystalline grains rather than on GBs in order to achieve maximal current density.


\section{Superconducting radio-frequency cavities}

Superconducting radio-frequency (SRF) cavities are essential components in particle accelerators due to the extremely low power loss under radio-frequency (RF) electromagnetic fields. The quality factor (Q) and the accelerating gradient (${E_{acc}}$) are two key parameters for SRF cavities \cite{SRF1}. Q describes the efficiency of the SRF cavity, which is determined by the surface resistance (${R_{s}}$). Substantial evidence points to the fact that the surface resistance results from the vortex penetrations and vortex motions in SRF cavities. The effect of vortices on the surface residual resistance is particularly important, as even at densities of the trapped flux as low as that of the geomagnetic field can lead to ${R_{res} \sim}$ 1 n${\Omega}$ in magnetically shielded high-purity niobium cavity at 2 K and 2 GHz. Several reports show that grain boundaries can give rise to a strong nonlinearity of the electromagnetic response of polycrystalline superconductors, which is crucial for Nb${_3}$Sn thin-film coatings of high-Q resonator cavities in particle accelerators \cite{SRF2,SRF3,SRF4,SRF5}.

Another key parameter of SRF cavities is ${E_{acc}}$ which determines how much energy can be transferred to a charge particle beam, which in turn is proportional to the peak magnetic field parallel to the cavity surface \cite{SRF1}. The superheating field $H_{\rm sh}$ is the peak magnetic field that the SRF cavities can withstand, which is estimated by indirect experiments based on detecting the onset of the nonlinearity of initial magnetization in superconductor \cite{SRF6,SRF7,SRF8} and by direct experiments of imaging vortices with a scanning Hall probe microscope \cite{SRF9}.

Thus, optimizing both the quality factor and the accelerating gradient, requires a profound understanding on the physics behind the superheating field and vortex dynamics of SRF cavities. The increasing interests in understanding the vortex nucleation and vortex dynamics of SRF cavities under AC magnetic fields have motivated the applications of TDGL simulations to the particular case of SRF cavities.

In 2017, Liarte \textit{et al.} \cite{SRF10} used TDGL theory to develop a 2D system with external magnetic field perpendicular to the 2D plane. This model allows one to investigate how the disorder, inhomogeneities, and material anisotropy influence the vortex nucleation under a parallel field and help to identify the regimes or configurations that can have negative consequences for cavity performance. It provides valuable information for understanding the effects of trapped vortices on the residual magnetic flux in SRF cavities.

Later on, the superheating field of a SRF cavity under an AC magnetic field and an AC current has been studied \cite{SRF11,SRF12}. Sheikhzada \textit{et al.} calculated the dynamic depairing current density and calculated the current-dependent kinetic inductance both in equilibrium and non-equilibrium states near the critical temperature $T{_c}$ by solving the TDGL equations. They found that an ac current density produces multiple harmonics of the electric field, with the amplitudes of the higher-order harmonics diminishing as ${\tau_E}$ increases. Their calculation results first demonstrated that the dynamic superheating field at $T$ ${\approx}$ $T_{\rm c}$ can exceed the static $H_{\rm sh}$ by 41{\%} \cite{Gurevich11}.

Real SRF resonant cavities inevitably contain impurities and defects \cite{50,SRF12.5,SRF13,SRF14,SRF15,SRF16}. The effect of surface roughness or point-like defects and grain boundaries on vortex penetration and heating dissipation in SRF cavities has been explored by TDGL equations. It was found that the surface defects and inhomogeneities will decrease the $H_{\rm sh}$ and cause the degradation of SRF properties. Pack \textit{et al.} \cite{50} simulated the nucleation of vortices in the presence of both geometric and material inhomogeneities by solving TDGL equations with finite-element method. In the two-dimensional case, they defined two geometries: an infinite cylinder and a thin film. In the 3D case they considered a rectangular box cut out of a thin film. The numerical results showed that even a small amount of disorder localizes the critical mode and can have a significant reduction in the effective superheating field for a particular sample. However, their results using the 3D model indicated that when defects are perpendicular to the magnetic field, the superheating field is effectively raised. Oripov \textit{et al.} \cite{SRF14} solved the TGDL equations with COMSOL software for a semi-infinite superconductor with point-like defects under a spatially non-uniform applied RF magnetic field. These simulations have proven very useful in understanding the measured third-harmonic response of Nb materials, subjected to intense localized RF magnetic fields. The simulations also indicate that vortex semi-loops are attracted towards the defect location and are distorted in shape. It is plausible that vortex semi-loops are one of the key sources of dissipation inside an SRF cavity at high operating power. 

\begin{figure}
	\centering
	\includegraphics*[width=0.9\linewidth,angle=0]{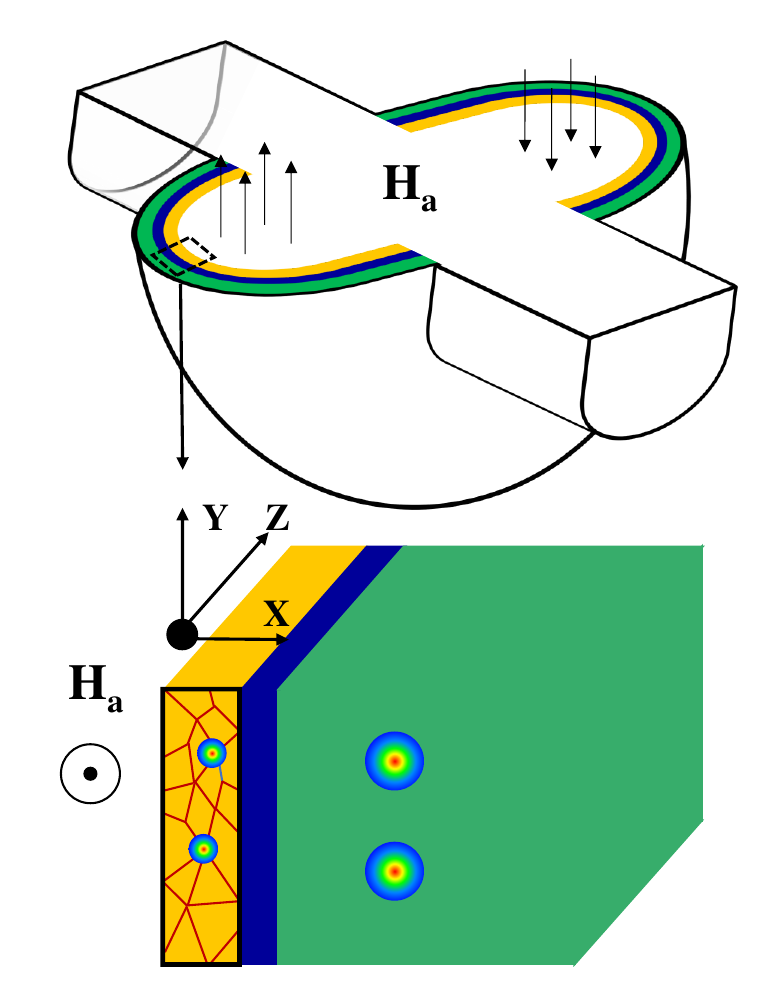}
	\caption{A schematic view of the S-I-S multilayer structure. The S-I-S structure is infinite along the $z$ axis and all the layers are parallel to the $y-z$ plane in the numerical model. The external magnetic field ${H_a}$ is along the $z$ axis. The material for the coating layer is ${\rm{Nb_3Sn}}$ (or ${\rm{NbN}}$) and the material for the bulk superconductor is Nb. Adapted from Ref. \cite{SRF21}}
	\label{SRF-1}
\end{figure}

Carlson \textit{et al.} \cite{SRF15} studied the mechanism of effects of GBs on magnetic vortex nucleations in Nb${_3}$Sn superconducting RF cavities and estimated the losses due to vortices filling grain boundaries. The numerical model consists of a film which lies perpendicular to an applied magnetic field $H_a$. Their calculations indicate that the grain boundaries act as both nucleation sites for vortex penetration and pinning sites for vortices after nucleation, depending on the magnitude of the applied field. Vortices nucleate and annihilate once per RF cycle. They estimated the losses associated to this process. It indicated that as long as vortices do not penetrate the bulk grain, losses are localized near the grain boundary, which does not lead to a global quench. Wang \cite{SRF16} performed TDGL simulations to better understand the measured third harmonic response, and compare these to the results obtained in Nb/Cu films. The third harmonic response of RF vortex nucleation caused by grain boundaries agree qualitatively well with the experimental data. Moreover, the density of surface defects that nucleate RF vortices, and how deep an RF vortex travels through these surface defects, can be extracted qualitatively from the measurements. They showed that the HiPIMS 75 V bias Nb/Cu sample is the best sample for SRF applications from the point of view of these two properties.

TDGL theory has also been used to study how to optimize the SRF properties. Sitaraman \textit{et al.} \cite{SRF17} used TDGL equations to extend the theory of hydride dissipation to sub-surface hydrides and calculate the high-field Q-slope (HFQS) caused by surface hydrides. They showed that sub-surface hydrides cause HFQS behavior. They found that the abrupt onset of HFQS is due to a transition from Meissner state to mix state and it is not necessary to entirely remove hydrides in order to eliminate the HFQS. It is only necessary to eliminate large hydrides which trigger the transition into the vortex state significantly below the niobium superheating field. Thus, creating more hydride nucleation sites near the surface can have a beneficial effect by decreasing the characteristic size of hydrides. 

The present record field of conventional Nb cavities \cite{SRF18,SRF19} is close to the theoretical field limit ($H_{\rm sh}$). Yet, an alternative approach to further improve the performance of SRF cavities has been proposed by Gurevich \cite{SRF20}, by using S-I-S multilayer structures (see Figure \ref{SRF-1}). As shown in Fig.~\ref{SRF-1}, a 2D numerical model was proposed in Ref. \cite{SRF21} for this multilayer structure, where a Nb${_3}$Sn-insulator-Nb structure is assumed to be infinite along the $z$-axis.

The TDGL equations for the S-I-S multilayer structure can be written as

\begin{equation}
	\label{GL11}
	\begin{split}
		\frac{\hbar^2}{2m_{s1}D_1}{\left(\partial_t+i\frac{e_s}\hbar\varphi\right)}\psi&=\frac{\hbar^2}{2m_{s1}}{\left(\nabla-i\frac{e_s}\hbar\mathbf{A}\right)^2}\psi\\
		&+{\left|a_1(T)\right|}\psi-b_1{\left|\psi\right|^2}\psi 
	\end{split}
\end{equation}

\begin{equation}
	\label{GL21}
	\begin{split}
		\frac1{\mu_0}\nabla\times\left(\nabla\times\mathbf{A}-\mu_0\mathbf{H}\right)&=\frac{\hbar e_s}{2m_{s1}i}\left(\psi^*\nabla\psi-\psi\nabla\psi^*\right)\\
		&-\frac{e_s^2}{m_{s1}}|\psi|^2\mathbf{A}+\sigma_1\left(-\nabla\varphi-\partial_t\mathbf{A}\right)
	\end{split}
\end{equation}

\begin{equation}
	\label{GL12}
	\begin{split}
		\frac{\hbar^2}{2m_{s2}D_2}{\left(\partial_t+i\frac{e_s}\hbar\varphi\right)}\psi&=\frac{\hbar^2}{2m_{s2}}{\left(\nabla-i\frac{e_s}\hbar\mathbf{A}\right)^2}\psi\\
		&+{\left|a_2(T)\right|}\psi-b_2{\left|\psi\right|^2}\psi 
	\end{split}
\end{equation}

\begin{equation}
	\label{GL22}
	\begin{split}
		\frac1{\mu_0}\nabla\times\left(\nabla\times\mathbf{A}-\mu_0\mathbf{H}\right)&=\frac{\hbar e_s}{2m_{s2}i}\left(\psi^*\nabla\psi-\psi\nabla\psi^*\right)\\
		&-\frac{e_s^2}{m_{s2}}|\psi|^2\mathbf{A}+\sigma_2\left(-\nabla\varphi-\partial_t\mathbf{A}\right)
	\end{split}
\end{equation}

\noindent where subscripts ‘1’ and ‘2’ denote Nb${_3}$Sn and Nb, respectively. ${D_1}$, ${a_1}$, ${b_1}$, ${D_2}$, ${a_2}$ and ${b_2}$ are phenomenological constants. Here, ${\varphi}$ is the electric scalar potential, which is included to retain the gauge invariance of the equation; $m_{s1}$ and $m_{s2}$ represent effective masses of the Cooper pairs in Nb$_{3}$Sn and Nb, respectively; $e_s$ is the effective charge; ${\hbar}$ is the Planck constant; ${\mu_0}$ is vacuum permeability; and $H$ is the external magnetic field. 

The dimensionless process of TDGL-I equations are as follows. Lengths are scaled by the coherence length ${\xi=\hbar/\sqrt{2m_{s1}a_1(0)}}$, time by Ginzburg–Landau relaxation time ${\tau=\xi^2/D_1}$, ${\psi}$ by ${\psi_0=\sqrt{|a_1(0)|/b_1}}$. Note that ${\xi{e_s}/\hbar=1/A_0}$, ${\tau{e_s}/\hbar=1/\varphi}$ and ${\varphi(\mathbf{r})=0}$ at all times, the equations (\ref{GL11}) and (\ref{GL12}) can be simplified to

\begin{equation}
	\partial_t\psi=\left(\nabla-i\mathbf{A}\right)^2\psi+\left(1-T_1\right)\psi-\left|\psi\right|^2\psi 
\end{equation}

\begin{equation}
	\partial_t\psi=\frac{D_2}{D_1}(\nabla-i\mathbf{A})^2\psi+\frac{m_{s2}D_2a_2(0)}{m_{s1}D_1a_1(0)}(1-T_1)\psi-\frac{m_{s2}D_2b_2}{m_{s1}D_1b_1}|\psi|^2\psi 
\end{equation}

\noindent where ${T_1=T/T_{c1}}$ and ${T_2=T/T_{c2}}$. $T$, ${T_1}$ and ${T_2}$ are working temperature, critical temperature for ${\rm{Nb_3Sn}}$ and Nb, respectively. 

The magnetic field is scaled by ${H_{c2}=\Phi_0/2\pi\xi^2}$. The resistivities ${\sigma_1}$ and ${\sigma_2}$  are scaled by the normal resistivity ${\sigma_0=1/\kappa_1^2D_1\mu_0}$ and we set ${\sigma_1/\sigma_0}$. TDGL-II equations (\ref{GL21}) and (\ref{GL22}) can be simplified to 

\begin{equation}
	\label{GL101}
	\partial_{t} \mathbf{A}=\frac{1}{2 i} \cdot\left(\psi^{*} \nabla \psi-\psi \nabla \psi^{*}\right)-|\psi|^{2} \cdot \mathbf{A}-\kappa_{1}^{2} \cdot \nabla \times \nabla \times \mathbf{A}
\end{equation}

\begin{equation}
	\label{GL102}
	\begin{split}
		&\partial_{t} \mathbf{A}=\frac{1}{2 i} \cdot\frac{\kappa_1^2D_1m_{s2}}{\kappa_2^2D_2m_{s1}}\cdot\left(\psi^{*} \nabla \psi-\psi \nabla \psi^{*}\right)-\\
		&\frac{\kappa_1^2D_1m_{s2}}{\kappa_2^2D_2m_{s1}}\cdot|\psi|^{2} \cdot \mathbf{A}-\kappa_{1}^{2} \cdot \nabla \times \nabla \times \mathbf{A}
	\end{split}
\end{equation}
The effects of temperature and GBs are introduced in TDGL equations by the function $f({\textbf{r},t})$, which can be denoted as $1-T(\textbf{r},t)-g(\textbf{r})$. The temperature gradient is imposed through the temperature field $T(\textbf{r},t)$ that is a function of the distributed over space and time. The pinning effect on vortices mainly comes from the suppressed superconductivity at the GBs, which can be modulated by the $g(\textbf{r})$ function in the TDGL equation. We define the average superconducting electron density at GBs (${|\psi|_{GB}}$) as a parameter to quantify the suppression of superconductivity on GBs, which can be tuned by function $g(\textbf{r})$. A temperature-independent GL parameter $\kappa$ has been adopted in the above TDGL equations, i.e., the same functional temperature dependence of both coherence length and penetration depth. This is an acceptable approach for temperatures not too far from the critical temperature. However, for simulations intending to cover the whole temperature range $0<T<T_c$, there is experimental evidence demonstrating that $\kappa$ varies with temperature \cite{paskin1965temperature,p2012temperature}. This effect that can be included in the TDGL equations by including suitable temperature-dependent kernels \cite{52}. Furthermore, TDGL offer a large degree of flexibility to accommodate spatially dependent material parameters such as mean-free-path, conductivity, critical temperature, etc. A highly illustrative example showcasing the success of this approach in direct comparison con experimental results can be found in Ref. \cite{spatially}. 

\begin{figure}
	\centering
	\includegraphics*[width=1.0\linewidth,angle=0]{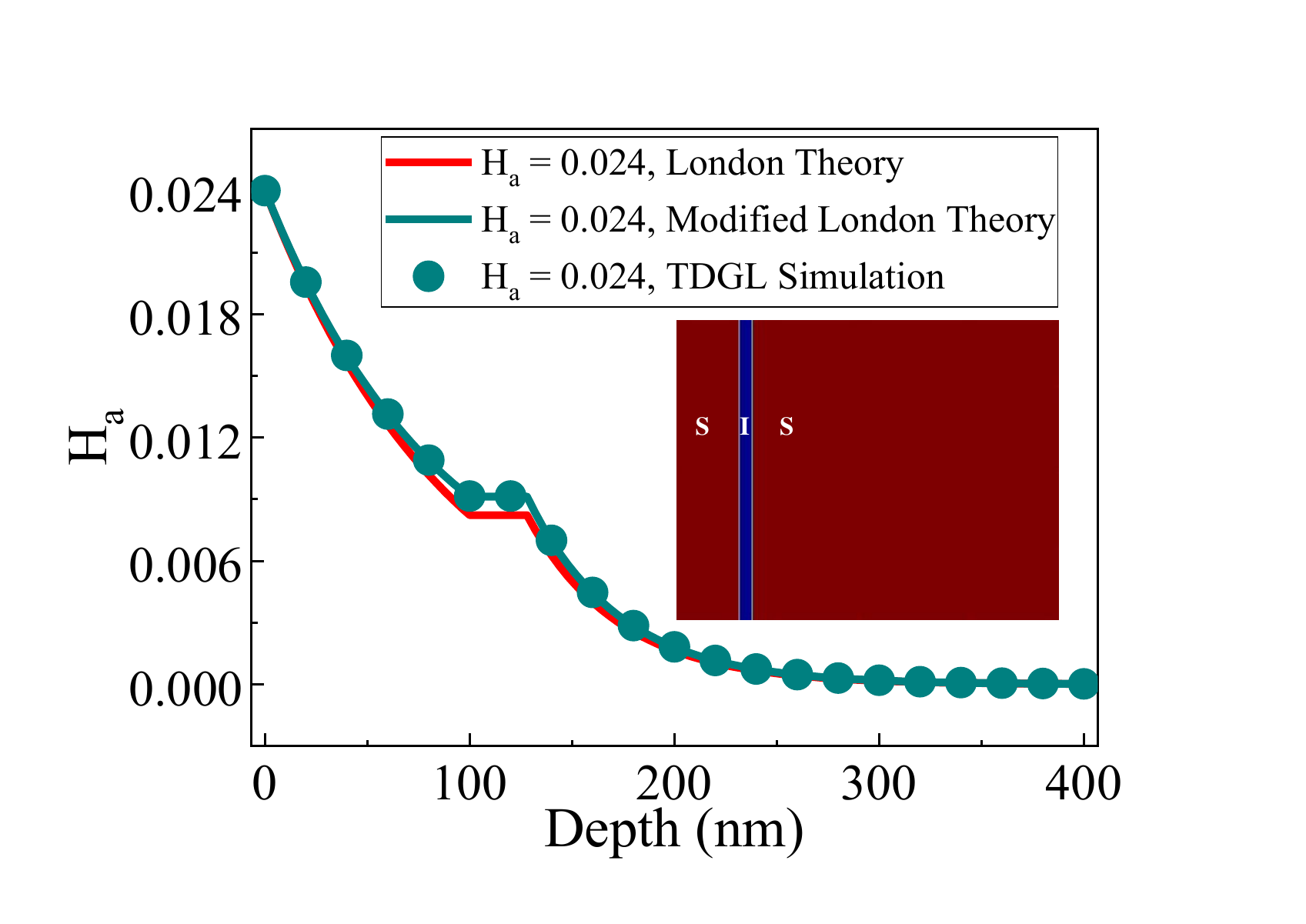}
	\caption{Magnetic field attenuation along the $x$-axis in the Nb${_3}$Sn${-}$insulator${-}$Nb multilayer structure (${\lambda_1}$ = 84 nm and ${\lambda_2}$ = 40 nm) calculated using TDGL theory and London theory at 0 K. Adapted from Ref. \cite{SRF21}.}
	\label{SRF-2}
\end{figure}

\begin{figure*}[!t]
	\centering
	\includegraphics*[width=1.0\linewidth,angle=0]{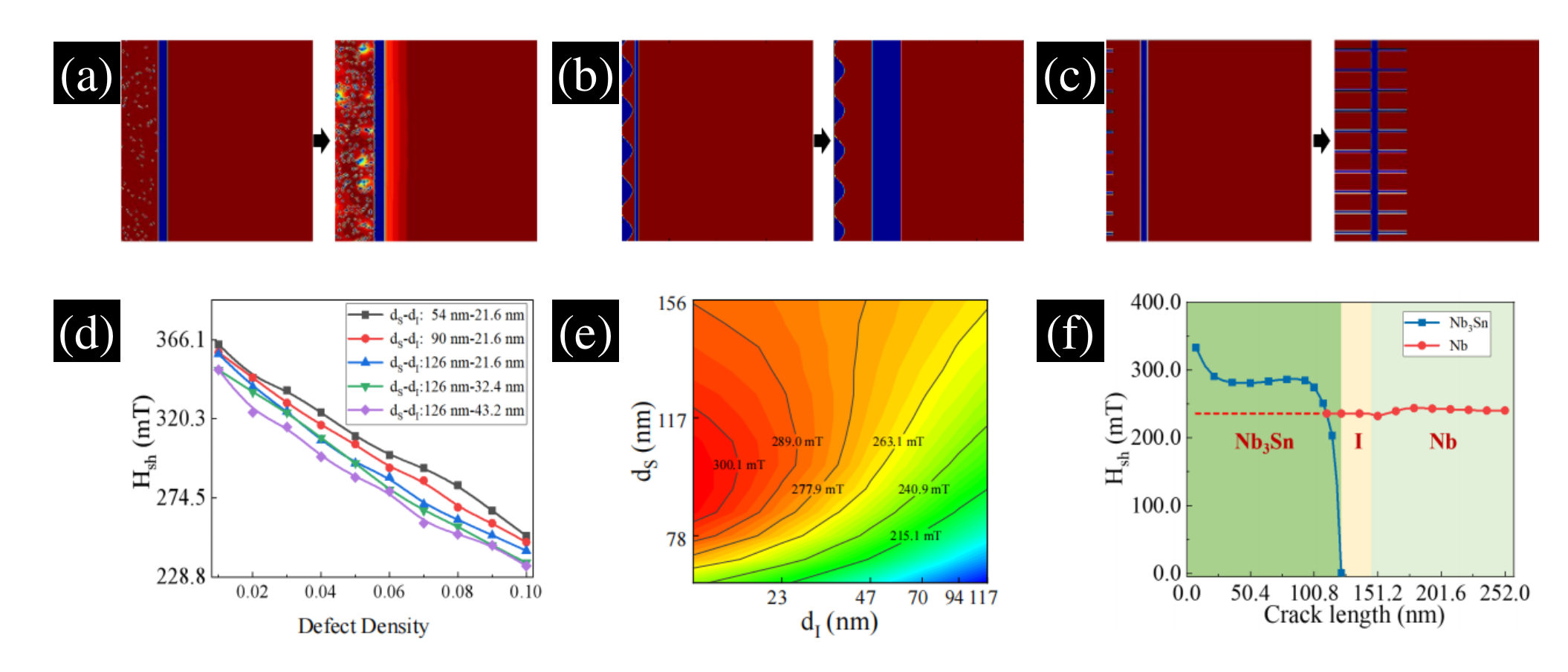}
	\caption{Schematic views of S-I-S cavity with (a) pointlike defects, (b) surface roughness, (c) cracks, and $H_{\rm sh}$ attenuation due to (d) pointlike defects, (e) surface roughness, (f) cracks. Adapted from Ref. \cite{SRF21}}
	\label{SRF-3}
\end{figure*}

\begin{figure*}[!t]
	\centering
	\includegraphics*[width=0.85\linewidth,angle=0]{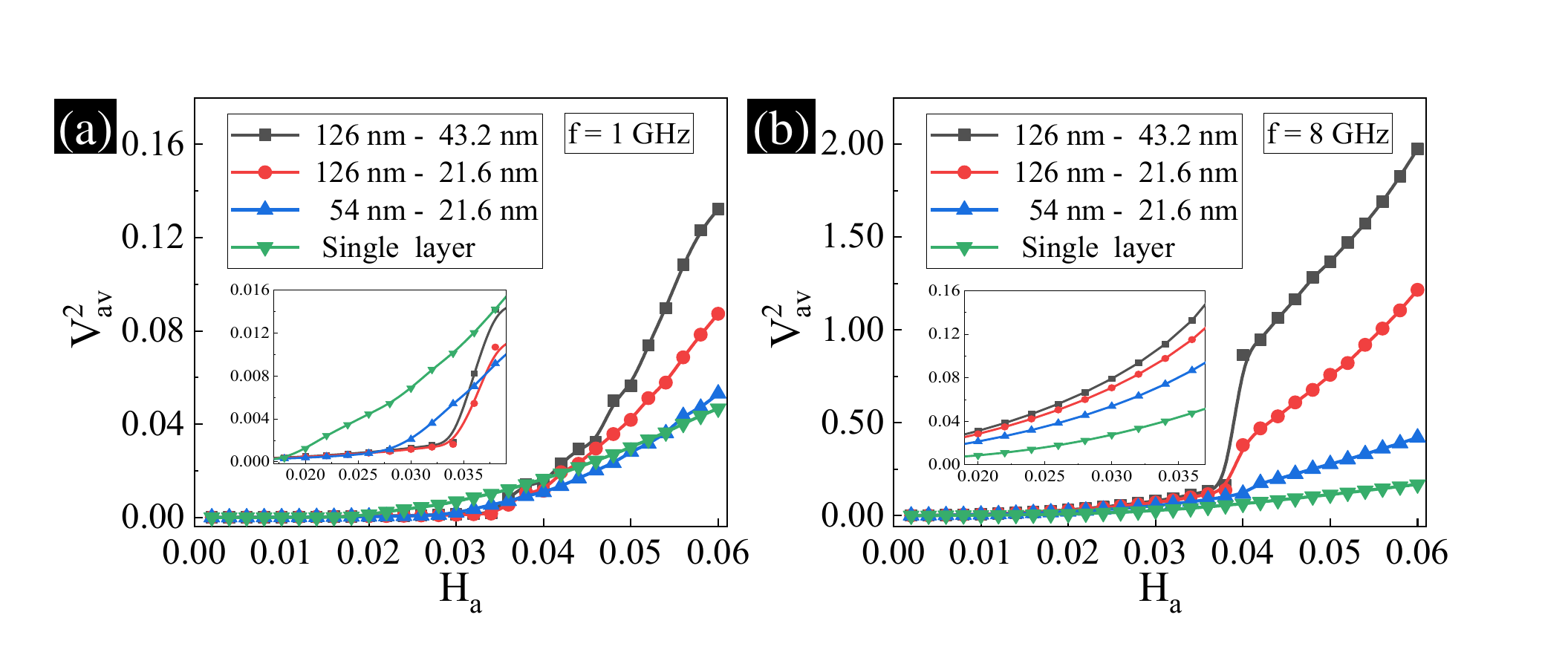}
	\caption{The variations of the voltage squared averaged over time and depth (${V^2_{av}}$) with increasing amplitude of RF magnetic field in SIS multilayer and single Nb structures at 1GHz (a) and 8GHz (b). Adapted from Ref. \cite{SRF21}.}
	\label{SRF-4}
\end{figure*}

 Experiments and numerical simulations clearly show that quickly cooling the sample can freeze the dynamic phase in the presence of a bias current and applied field \cite{Formation}. In order to consider the temperature variations induced by dissipated energy, the thermal diffusion equation (\ref{Heat}) is used to obtain the temperature field, where ${\lambda(T)}$ is the temperature-dependent thermal diffusion coefficient, and $C(T)$ is the temperature-dependent volume-specific heat. \textit{W} represents the heat source induced by vortex motion. The first term of equation (\ref{WWWW}) accounts for energy dissipation due to induced electric fields, and the second term is related to the relaxation of the order parameter. Considering that the contribution of the order parameter relaxation to the heat source can be neglected, we focused on the first term. The finite difference method(FDM) algorithm are employed to solve the heat diffusion equation with open boundary conditions along $x$ direction and periodic boundary condition along $y$ direction, respectively. 

\begin{equation}
	\nabla(\lambda(T)\nabla T)-C(T)\frac{\partial T}{\partial t}+W_s=0
	\label{Heat}
\end{equation}

\begin{equation}
	W=\left(\frac{\partial\mathbf{A}}{\partial t}\right)^2+\frac{2u}{\sqrt{1+\gamma^2|\psi|^2}}\left[\left|\frac{\partial\psi}{\partial t}\right|^2+\frac{\gamma^2}4\left(\frac{\partial|\psi|^2}{\partial t}\right)^2\right]
	\label{WWWW}
\end{equation}

\begin{figure*}[!t]
	\centering
	\includegraphics*[width=0.8\linewidth,angle=0]{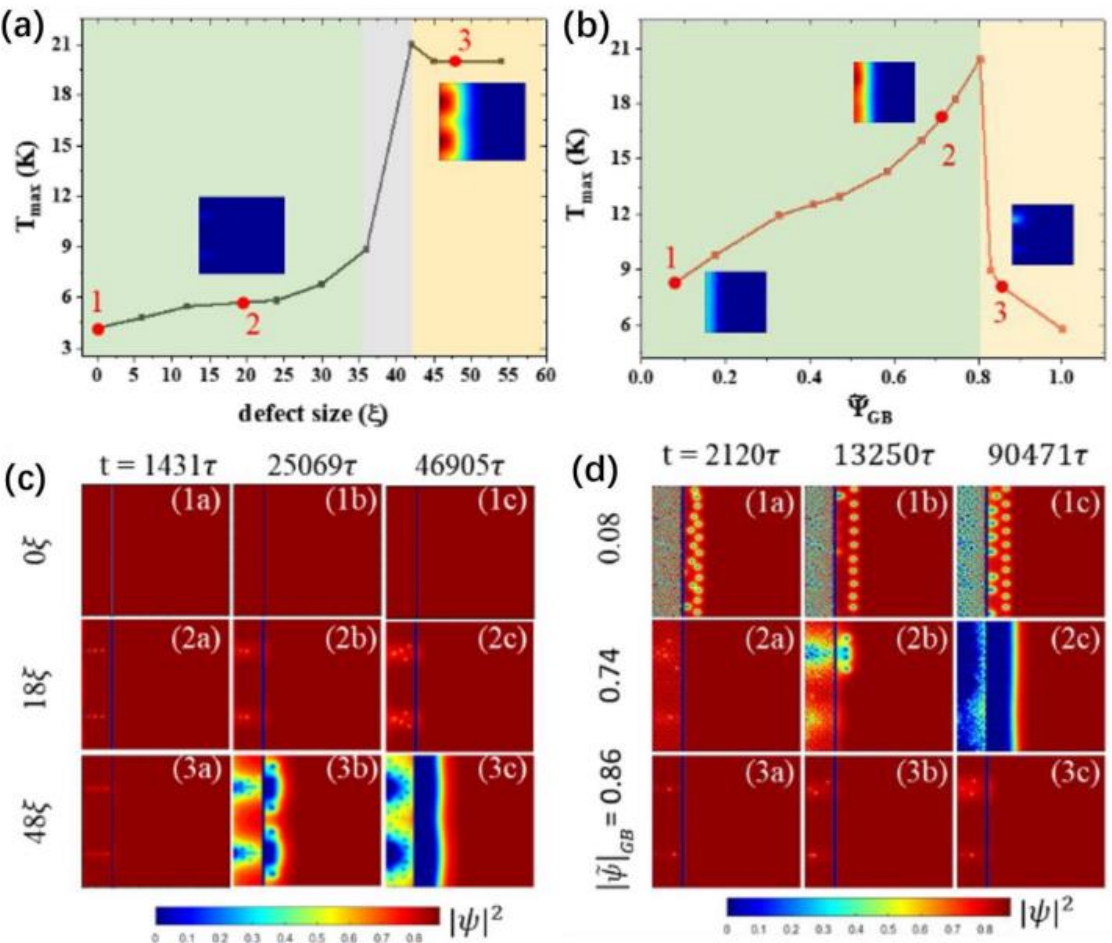}
	\caption{(a-b) Variations of maximum temperature with defect size and superconductivity of GBs $|\tilde{\psi|}_{GB}$ in S-I-S multilayer structures. Contour plots of Cooper pair density in S-I-S multilayer structures with/without surface defects (c) and with GBs of different $|\tilde{\psi|}_{GB}$ (d). Adapted from Ref. \cite{SRF34}}
	\label{SRF-5}
\end{figure*}

\begin{figure}[!t]
	\centering
	\includegraphics*[width=1.0\linewidth,angle=0]{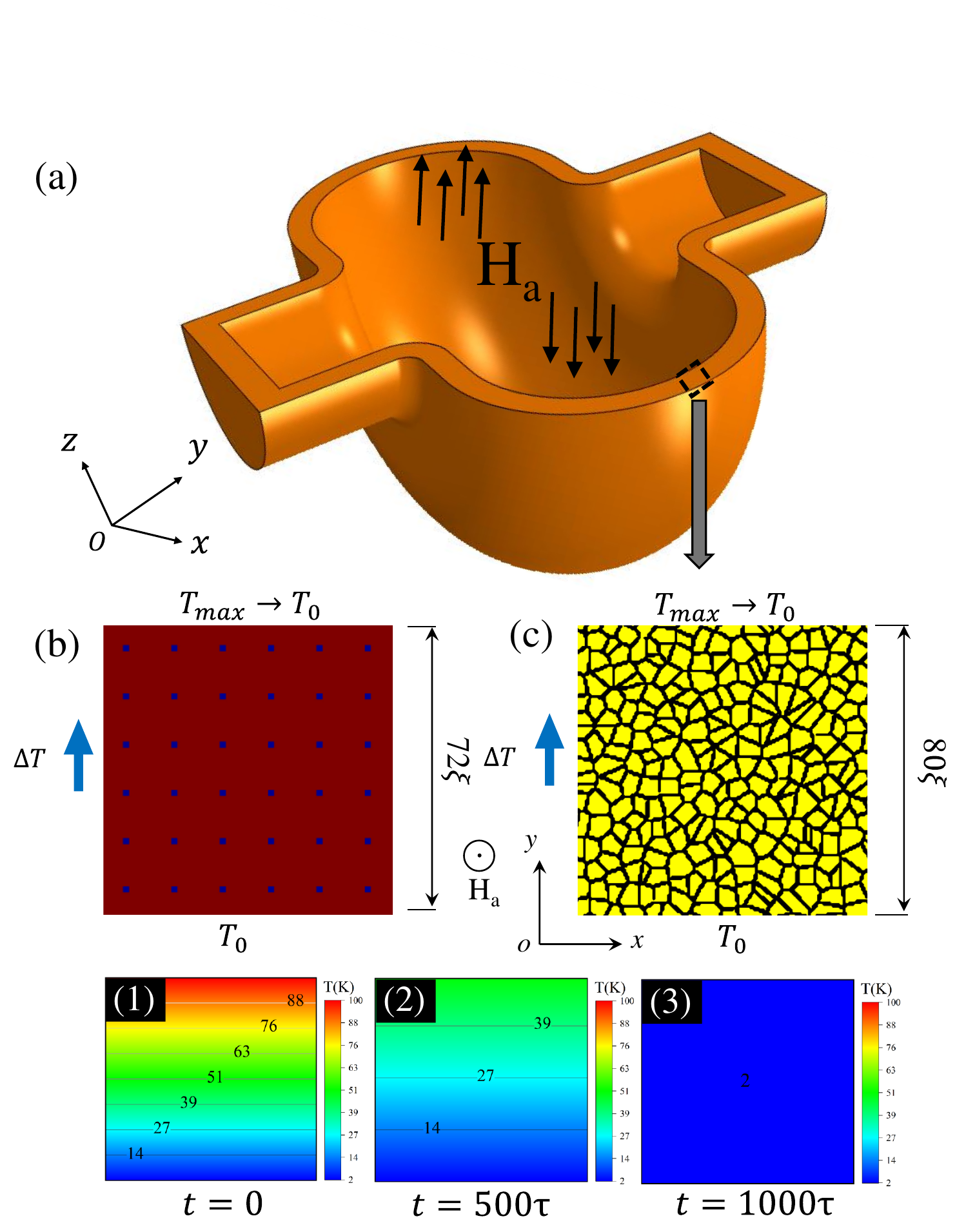}
	\caption{Schematic of SRF cavity structure (a) and numerical models with array of antidots (b) and GBs (c). Panels 1-3 show the variations of temperature gradient with time used in numerical simulations. Adapted from Refs. from \cite{SRF32} and \cite{SRF33}.}
	\label{SRF-6}
\end{figure}

Vodolazov \textit{et al.} \cite{SRF21.1,SRF21.2,SRF21.3} investigated the dependence of the critical current on magnetic field, photon detection, and photon-triggered instability in a superconducting strip, based on coupled TDGL and heat diffusion equations. Elmurodov \textit{et al.} \cite{SRF21.4} coupled the TDGL equations to heat diffusion and investigated the phase-slip phenomena in one dimensional (1D) NbN systems. Berdiyorov \textit{et al.} \cite{58} study a strip with holes by thermal-magnetic coupling and reveal the motion of current-excited vortices as an origin for the magnetoresistance oscillations in mesoscopic superconductors. Jing \textit{et al.} \cite{SRF21.6,SRF21.7} consider the thermal effect on the vortex dynamics of superconducting thin films. Antonio \textit{et al.} \cite{SRF21.8} show that a microwave closed to the excitation frequency of Pb can effectively inhibit flux avalanches in Pb superconducting films when approaching the critical temperature or magnetic field.  

We consider a Nb${_3}$Sn-insulator-Nb multilayer structure with 720 nm × 720 nm in the $x$ $–$ $y$ plane with infinite length and magnetic field along the $z$ $-$ axis. The size of simulated region is thus very small compared with that corresponding to real cavities. However, the fact that the magnetic field decreases within a short distance of about hundreds of nm and that SRF cavities exhibit rotation symmetry, permits us to faithfully describe the physical phenomena using the dimensions used in the simulation. Similarly, it is a common practice to experimentally determine the critical field and performance of the cavities from samples much smaller than the real cavity. Real cavities are made on oxygen-free copper (mm thickness) coated with a Nb layer (µm thickness).

Figure \ref{SRF-2} shows the magnetic field distributions of the SIS multilayer structure at $H_{\rm a}$ = 0.024. It can be seen from the magnetic field attenuation that our numerical results match the theoretical formulas quite well for the Nb${_3}$Sn–insulator–Nb multilayer structure.

The surface defects and surface roughness (Figure \ref{SRF-3}) in the S-I-S structure under quasi-static magnetic field not only has a significant impact on the superheating field, but also causes difficulties on the optimization strategy based on S-I-S structure. Although the number of pointlike defects increases with the thickness of the superconducting layer ($dS$), the decreasing slope of superheating field of the resonant cavity in S-I-S structures with density of defects seems to be independent on the thicknesses of superconducting layer and insulator layer. 

In order to describe the main characteristic of complex surface topographies of real materials, we choose a sine function to mimic the surface roughness (Figure \ref{SRF-3}). One can observe that $H_{\rm sh}$ is significantly reduced due to the surface roughness. This may be attributed to the fact that surface roughness changes the effective thickness of the coating layer. 

Fig. ~\ref{SRF-3}(c,f) show the variations of $H_{\rm sh}$ of the S-I-S multilayer structure ($dS = 126$ nm and $dI = 21.6$ nm) with length of cracks. Note that $H_{\rm sh}$ of the coating layer (blue solid line) decreases rapidly with the length of cracks. However, it is interesting to note that $H_{sh}$ decreases very slightly and even increases a little when the crack length is in the range from 25 nm to 100 nm.

The S-I-S structure cavity is generally operated under magnetic field with radio frequency. When the SRF cavity is exposed to a magnetic field with small amplitude and low frequency radio frequency, the induced voltage of the S-I-S structure resonant cavity is lower than that of the single-layer Nb structure resonant cavity (see figure \ref{SRF-4}). However, when the resonant cavity is exposed to a magnetic field with large amplitude and high frequency radio frequency, the induced voltage of the S-I-S structure resonant cavity is greater than that of the single-layer Nb structure resonant cavity. Moreover, the optimal superconducting layer thickness and maximum insulation layer thickness under AC fields are inconsistent with those under quasi-static magnetic fields.  Note that figure \ref{SRF-4} reveals that the superheating field tends to decrease as frequency increases. The simulated results indicate that the effects of the frequency, amplitude, surface roughness, and defects of the RF magnetic field should be considered when evaluating the actual performance of the resonant cavity of the S-I-S multilayer structure. Further research on the optimal superconducting layer thickness and optimal insulation layer thickness under alternating fields is also needed.

Pathirana \textit{et al.} calculated the DC superheating field for superconductors with an inhomogeneous impurity concentration or with S-I-S structures by TDGL \cite{SRF22}. The rapid vortex motion accompanied by high energy dissipation can cause thermomagnetic instabilities in SRF cavities. The local thermomagnetic instability can lead to quenches due to rapid local temperature rise, and even induce damage in the superconducting cavity. In addition, microscopic defects on the inner surface of the superconducting cavity and grain boundaries in the Nb$_3$Sn material are crucial on the vortex dynamics. Therefore, it is necessary to explore the vortex dynamics by coupling the TDGL equations and thermal diffusion equation \cite{Eley} in order to accurately simulate the performance of the SRF cavity.

Wang \textit{et al.} \cite{SRF34} explored the effects of tiny surface defects on the thermomagnetic instabilities of the superconducting cavity. As shown in Fig. ~\ref{SRF-5}(a, c), the S-I-S structure can remain in the Meissner state, whereas surface defects lead to vortex penetration, which causes local rapid increase of temperature (but not quench yet). particularly, when the defect-size exceeds a certain value, it causes global quench. The thermomagnetic instabilities also depends on the amplitude and frequency of magnetic field.

For the case of Nb${_3}$Sn layer with both GBs and small surface defects, as shown in Fig.~\ref{SRF-5}(b, d), the influence of GBs on thermomagnetic instabilities of the S-I-S multilayer structure is not monotonic. When the superconductivity of GBs is weak, vortices mainly enter the superconductor through GBs, which leads to small energy dissipation. With increasing the superconductivity of GBs, more vortices penetrate into the superconductor and interstitial vortices inside crystalline grain can be observed, which can cause global quenching. However, the superconductivity of GBs exceeds a certain value, the maximum temperature decreases. In this case, vortices mainly nucleate at surface defects and the energy dissipation by vortex motion is much smaller than aforementioned cases.

It has been found that the number of vortices trapped nearby surface of SRF cavity is very strongly dependent on the temperature gradient. \cite{SRF24,SRF25,SRF26,SRF27}. In addition, ultra-high-vacuum (UHV) furnace treatment not only changes the micro-structure inside the cavity such as grain size and dislocation content \cite{SRF28,SRF30}, but also has a significant influence on flux expulsion. A large number of experiments have been done to explore the effective of flux expulsion by space temperature gradients (temperature brooms) \cite{SRF25,SRF30,SRF31}. Due to pinning effect of numerous defects in pure niobium cavities on vortices, it is difficult to expel vortices from the sample, which reduces the quality factor $Q_{0}$ of the SRF cavities. It was found that cooling the superconducting cavities with a large spatial temperature gradient is an effective way to expel the trapped magnetic flux. Based on TDGL theory, as shown in Fig. ~\ref{SRF-6}, Li \textit{et al.} \cite{SRF32} and He \textit{et al.} \cite{SRF33} investigated the effect of linear temperature gradient on magnetic flux expulsion in pure Nb and Nb${_3}$Sn SRF cavities with antidots and GBs. The initial maximum temperature $T_{max}$ (greater than $T_c$) is decreased to working temperature $T_{0}= 2$ K with time in the simulations. They demonstrated that the temperature broom can remove the vortices from clean sample more easily than from the samples with defects. Moreover, larger spatial temperature gradient can sweep more vortices out of the samples. 

As for the samples with GBs, the effects of grain size, pinning potential of GBs, and rates of temperature gradient with time on flux expulsion were theoretically investigated. Similar to the case of antidots, larger temperature gradient is more effective to sweep the vortices, which is in excellent agreement with the experimental results \cite{SRF24,SRF30,SRF31}. For strong superconductivity of GBs (large $|\psi|_{GB}$), the expelling efficiency of temperature gradient on vortices is a monotonic function of grain size. However, for weak superconductivity of GBs (small $|\psi|_{GB}$), the expelling efficiency of temperature gradient on vortices first increases and then decreases with increasing the grain size.

\section{Conclusions}
Due to their unique electromagnetic properties, superconductors continue to attract considerable academic interest not only because of their rich underlying physical mechanisms, but also motivated by the fast developing body of applications in superconducting electronics and high current and/or high magnetic fields. Even though there is growing evidence suggesting that new materials with superconducting critical temperatures will be able to reach high enough values requiring conventional cooling apparatuses, as of today technological relevant superconductors still require rather costly refrigeration systems. In this context, the invaluable help and contribution of numerical tools become commendable. In this quest, TDGL equations provide an elegant and powerful tool to investigate the electromagnetic properties of superconductors for which the macroscopic properties are ultimately dictated by the microscopic behavior of vortices. By means of high-performance parallel computation with graphics processing units (GPUs), large-scale TDGL simulations can be employed to deal with samples having realistic pinning landscapes and large sizes thus reconnecting with alternative numerical mean-value approaches. Three case studies were presented and discussed, namely (i) superconducting vortex rectifications and/or superconducting diode effect, (ii) critical current density of vortex pinning/depinning in the presence of various pinning landscapes, and (iii) vortex penetrations in SRF cavities. Hand by hand with the continuous development of computing power along with more efficient algorithms, TDGL simulations are progressively approaching the complexity inherent to real devices with 3D architectures, large sizes, and multi-physics (thermal equations, strain, etc.) which will eventually lead to the ultimate goal of improving the electromagnetic properties of superconductors by design.

\acknowledgements{We acknowledge support by the National Natural Science Foundation of China (Grant Nos. 12372210 and 11972298).}

\bibliographystyle{emsreport}
\bibliography{refs}



\authorinfo{author1}%
{Cun Xue}%
{is currently an associate professor of Department of Engineering Mechanics, at Northwestern Polytechnical University. He received the B.S. and Ph.D. degrees from Lanzhou University of solid mechanics, Lanzhou, China, in 2010 and 2016, respectively. He was a visiting student at KU Leuven Belgium during 2014-2015. His research interests mainly focus on multifield coupling problems of the applications of superconductors, including superconducting magnets, superconducting radio-frequency cavities, and superconducting devices.}%
{xuecun@nwpu.edu.cn}

\authorinfo{author2}%
{Qing-Yu Wang}%
{received the B.S. degree from Lanzhou University, China, in 2019. He is currently a Ph.D. student in the School of Aeronautics, Northwestern  Polytechnical University. His current research interests include thermomagnetic instability, vortex dynamics of superconductor under mechanical loading. }
{qywang@mail.nwpu.edu.cn}

\authorinfo{author3}%
{Han-Xi Ren}%
{received the B.S. degree from Northwestern Polytechnical University, China, in 2023. He is currently a Ph.D. student in the School of Aeronautics, Northwestern  Polytechnical University. His current research interests include multi-field coupling of high-field Nb$_3$Sn superconducting magnet coils and optimization of critical properties of superconducting wires. }%
{renhanxi@mail.nwpu.edu.cn}

\authorinfo{author4}%
{An He}%
{is an associate professor at the School of Science, Chang 'an University. She received her B.S. and Ph.D. degrees from Lanzhou University in 2011 and 2016, respectively. She was a visiting student at University of Antwerp Belgium during 2014-2015.  She mainly engaged in multi-field coupling performance analysis of superconducting materials.}%
{hean@chd.edu.cn}

\authorinfo{author5}%
{Alejandro Silhanek}%
{is a professor of Department of Physics, at the University of Liège, Belgium. He received the Ph.D. degree in Physics from Instituto Balseiro, Bariloche (Argentina) in 2001. He was awarded Directors funded Postdoctoral fellow, Los Alamos National Laboratory (USA) from 2004 to 2006 and Postdoctoral Researcher at the KU Leuven Belgium from 2006 to 2011. The scientific activities of his group are mainly within the field of experimental mesoscopic physics, nanoscience and nanotechnology, magnetism, superconductivity, electromigration, low -frequency metamaterials and quantum transport.}%
{asilhanek@uliege.be}


\end{document}